\documentclass[aps,twocolumn,epsfig,graphics,showpacs,floatfix,mathbbm]{revtex4}

\usepackage{amsmath,amsfonts,amssymb,graphics,graphicx,epsfig,color,times,bbm}
\usepackage{amsthm}
\usepackage{psfrag}

\newcommand{\me}{\mathrm{e}}

\newcommand{\md}{\mathrm{d}}

\newcommand{\zz}{\mathbbm{Z}}
\newcommand{\nn}{\mathbbm{N}}
\newcommand{\rr}{\mathbbm{R}}
\newcommand{\id}{\mathbbm{1}}

\newtheorem{theorem}{Theorem}

\newtheorem{lemma}[theorem]{Lemma}

\newtheorem{corollary}[theorem]{Corollary}

\newcommand{\stale}{{\emptyset}}

\begin{document}

\title{Potential and limits to cluster state
quantum computing using probabilistic gates}

\author{D.\ Gross, K.\ Kieling, and J.\ Eisert}

\affiliation{
1 QOLS, Blackett Laboratory, 
Imperial College London,
Prince Consort Road, London SW7 2BW, UK\\
2 Institute for Mathematical Sciences, Imperial College London,
Prince's Gate, London SW7 2PG, UK
}
\date\today

\begin{abstract}
We establish bounds to the necessary 
resource consumption when building up cluster states
for one-way computing using probabilistic gates. 
Emphasis is put on state preparation with
linear optical gates, as the probabilistic character
is unavoidable here. 
We identify rigorous general bounds to the 
necessary consumption of initially available
maximally entangled pairs when building
up one-dimensional cluster states with 
individually acting linear 
optical quantum gates, entangled pairs 
and vacuum modes. 
As the known linear optics gates have a limited maximum 
success probability, as we show, this 
amounts to finding the optimal classical 
strategy of fusing pieces of 
linear cluster states. A formal
notion of classical configurations and  
strategies is introduced for probabilistic non-faulty
gates. We study the asymptotic performance
of strategies that can be simply described, and 
prove ultimate bounds to the 
performance of the globally optimal strategy.
The arguments employ methods of 
random walks and convex optimization.
This optimal strategy is also the one that
requires the shortest storage time, and necessitates
the fewest invocations of probabilistic
gates. For 
two-dimensional cluster states, we find, for any
elementary  success probability, 
an essentially deterministic preparation of 
a cluster state with 
quadratic, hence optimal, asymptotic scaling in the use of
entangled pairs.
We also identify a percolation effect in state preparation,
in that from a threshold probability on, 
almost all preparations will be either 
successful or fail. We outline the implications on
linear optical architectures and 
fault-tolerant computations.
\end{abstract}

\maketitle

\section{Introduction}

Optical quantum systems offer a number of advantages
that render them suitable
for attempting to employ them in
architectures for a universal quantum
computer: decoherence is less of an issue
for photons compared to other physical systems, and
many of the tools necessary for quantum state manipulation are
readily available 
\cite{KLM,Rev,Reznik,Nielsen04,terry,Ralph05,PC}.
Also, the possibility of
distributed computation is an essentially built-in feature
\cite{PC,Dist,Dist2}.
Needless to say, any realization of a
medium-scale linear optical quantum computer
still constitutes an enormous challenge \cite{Computers}.
In addition to
the usual requirement of near-perfect hardware
components -- here, sources of
single photons or entangled pairs, linear optical networks, and
photon detectors -- one has to live with a further
difficulty inherent in this kind of architecture: 
due to the small success probability of elementary
gates, a very significant {\it overhead in
optical elements and additional photons}
is required to render the overall protocol 
near-deterministic.

Indeed, as there are no photon-photon interactions present in coherent
linear optics, all non-linearities have to be induced by 
means of measurements.
Hence, the probabilistic character is at the core of such 
schemes. It
was the very point of the celebrated work of Ref.\
\cite{KLM} that near-deterministic quantum computation is indeed
possible using quantum gates (here: non-linear sign shift gates) 
that operate with a very low probability of success: only
one quarter. Ironically, it turned out later that this value cannot be
improved at all within the setting of linear optics 
without feed-forward \cite{SP}. 
Essentially due to this small 
probability, an enormous overhead in resources in the full 
scheme involving
feed-forward is needed.

There is, fortunately, nevertheless 
room for a reduction of this overhead,
based on this seminal work.
Recent years saw a development reminiscent of a ``Moore's
law'', in that each year, a new scheme was suggested that reduced the
necessary resources by a large factor. In particular, the most
promising results have been achieved \cite{Reznik,Nielsen04,terry} by
abandoning the standard gate model of quantum computation
\cite{Computers} in favor of the measurement-based \emph{one-way
computer} \cite{Oneway}. Taking resource consumption as a benchmark,
the most recent schemes range more than 
two orders of magnitude below
the original proposal.  It is thus meaningful to ask: {\it How long
can this development be sustained? What are the ultimate limits to
overhead reduction for linear optics quantum computation?} The latter
question was one of the main motivations for our work.

The reader is urged to recall that a computation in the one-way model
proceeds in two steps. Firstly, a highly entangled \emph{cluster
state} \cite{Oneway,GS,LongGS,Exp} is built up. Secondly, local
measurements are performed on this state, the outcomes of which encode
the result of the computation. As the ability to perform local
measurements is part of the linear optical toolbox, the challenge lies
solely in realizing the first step. More specifically, one- and
two-dimensional cluster states can be built from EPR pairs \cite{EPR} using
probabilistic so-called \emph{fusion gates}.  In the light of this
framework, the question posed at the end of the last paragraph takes
on the form: \emph{measured in the number of required entangled pairs,
how efficiently can one prepare cluster states using probabilistic
fusion gates?} There have been several proposals along these lines
in recent years \cite{Reznik,Nielsen04,terry,BK,O2,DR,Letter}.

\begin{figure}
  \includegraphics{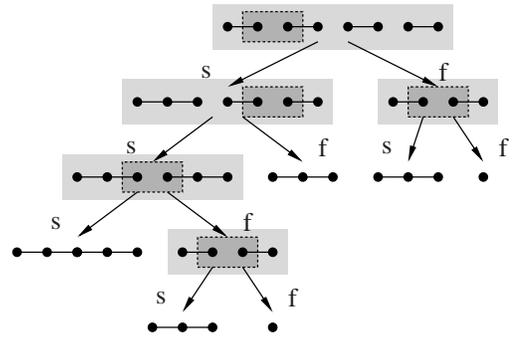}
  \caption{ An example of a tree of successive 
  configurations under application of a strategy. Light 
  boxes group configurations.
    We start with $N=4$. Dark boxes indicate
    where the strategy decided to apply a fusion gate. 
    Possible outcomes are success (to the left) 
		or failure (to the right), resulting in different possible future
		choices.  The expected length of the final chain is $\tilde
		Q_M(4)=Q(4)=13/8$. \label{fig:plethora} }
\end{figure}

It will be shown that the success probability of these
gates can not be pushed beyond the currently known 
value of one half.
Therefore, the only degree of freedom left in optimizing the process
lies in \emph{adopting an optimal classical control strategy, which
decides how the fusion gates are to be employed}. This endeavor is
greatly impeded by the gates' probabilistic nature: the number of
possible patterns of failure and success scales exponentially (see
Fig.~\ref{fig:plethora}) and hence deciding how to optimally react to
any of these situations constitutes a very hard problem indeed. 

Maybe surprisingly, we find that classical control has tremendous
implications concerning resource consumption (which seems particularly
relevant when building up structures that render a scheme eventually
fault-tolerant \cite{Trees,Trees2}): even when aiming for moderate
sized cluster states, one can easily reduce the required amount of
entangled pairs by an order of magnitude when adopting the appropriate
strategy. For the case of one-dimensional clusters, we identify a
limit to the improvement of resource consumption by very tightly
bounding from above the performance of any scheme
which makes use of EPR pairs, vacuum modes and two-qubit quantum
gates.  In the two-dimensional setup, we establish that cluster states
of size $n\times n$ can be prepared using $O(n^2)$ input pairs.

We aim at providing a comprehensive study of the potential and
limits to resource consumption for one-way computing, when the
elementary gates operate in a {\it non-faulty, but probabilistic
fashion}. As the work is phrased in terms of classical control
strategies, it applies equally to the linear optical setting as to
other architectures \cite{BK,DR,Port}, such as those involving {\it
matter qubits and light as an entangling bus} \cite{Plenio,BK}.
This work extends an earlier report (Ref.\ \cite{Letter})
where most ideas have already been sketched.

\section{Summary of results}

Although the topic and results have very practical implications on the
feasibility of linear optical one-way computing, we will have to
establish a rather formal and mathematical setting in order to obtain
rigorous results. To make these more accessible, we provide a short
summary:
\begin{itemize}
\item
	We introduce a formal framework of {\it classical strategies} for
	building up linear cluster states.  Linear cluster states can be
	pictured as {\it chains} of qubits, characterized by their length
	$l$ given in the number of edges.  Maximally entangled qubit pairs
	correspond to chains with a single edge.  By a {\it configuration}
	we mean a set of chains of specific individual lengths.  Type-I
	fusion \cite{terry} allows for operations involving end qubits of
	two pieces (lengths $l_1$ and $l_2$), resulting on success in a
	single piece of length $l_1+l_2$ or on failure in two pieces of
	length $l_1-1$ and $l_2-1$. The process starts with a collection of
	$N$ EPR pairs and ends when only a single piece is left.  A {\it
	strategy} decides which chains to fuse given a configuration.  It is
	assessed by the {\it expected length}, or \emph{quality} $\tilde
	Q(N)$ of the final cluster.  The vast majority of strategies allow
	for no simple description and can be specified solely by a ``lookup
	table'' listing all configurations with the respective proposed
	action. Since the number of configurations scales
	exponentially as a function of the total number of edges $N$, a
	single strategy is already an extremely complex object and any form
	of brute force optimization is completely out of reach.
	
\item After discussing the optimality of the
        primitive {\it elementary physical gates},
        operating with a success probability of $p_{\text s}=1/2$,
        we start by studying the performance of several {\it simple
        strategies}. In particular, we study strategies
        which we refer to as {\sc Modesty} and
        {\sc Greed}:
        \begin{eqnarray*}
        \text{\sc Greed}:&& \text{Always fuse the largest available}
        \nonumber\\
        &&\text{ linear cluster chains.}\\
        \text{\sc Modesty}:&& \text{Always fuse the smallest 
        available} \nonumber\\
        &&\text{ linear cluster chains}
        \end{eqnarray*}
        in a configuration.
        Also, we investigate the strategy {\sc Static}
        with a linear yield
        that minimizes the amount of sorting and 
        feed-forward.

\item We find that the choice of the
        classical strategy has a
        {\it major impact} on the resource consumption
        in the preparation of linear cluster states. When
        preparing cluster chains with an expected length of
        $40$, the number of required EPR pairs already differs
        by an order of magnitude when resorting to
        {\sc Modesty} as compared to {\sc Greed}.

\item  We provide an algorithm that symbolically
        identifies the \emph{globally optimal strategy}, which yields the
        longest average chain with a given number $N$ of
        initially available EPR pairs. This globally optimal
        strategy can be found with an effort of
        $O(|{\cal C}^{(N)} | (\log |{\cal C}^{(N)} |)^5)$. Here,
				$|{\cal C}^{(N)}|$ is the number of all configurations with up
				a total number of up to $N$ edges.

\item We find that {\sc Modesty} is almost globally
        optimal. For $N\leq 46$, the relative
        difference in the quality of the globally optimal strategy and
        {\sc Modesty} is less than $1.1\times 10^{-3}$.

\item Requiring significantly
        more formal effort, we provide
        fully {\it rigorous proofs} of {\it tight analytical
        upper bounds concerning the
        quality of the globally optimal strategy}.
        In particular, we find
        \begin{equation*}
                \tilde Q(N) \leq N/5 + 2.
        \end{equation*}
        That is, frankly, within the setting of linear optics,
        in the sense made precise below, one has to
        invest at least {\it five EPR pairs per
        average gain of one edge in the cluster state}. 

\item 
				A key step in the proof is the passage to a radically
				simplified model -- dubbed {\it razor model}. Here, cluster
				pieces are cut down to a maximal length of two. While this step
				reduces the size of the configuration space tremendously, it
				retains -- surprisingly -- essential features of the problem.
				The whole problem can then be related to a {\it random walk in
				a plane}~\cite{VK}, and finally, to a convex optimization problem~\cite{Convex}. This
				bound constitutes the central technical result.

\item The {\it razor model} also provides tools to
        get good
        numerical upper bounds with {\it polynomial effort} 
        in $N$.

\item Similarly, we find tight {\it lower bounds} for the
        quality, based on the symbolically available data for
        small values of $N$.

\item We show that the questions ({\it i}) 
``given some fixed number of input pairs, how long a single chain can
be obtained on average?'' and ({\it ii}) ``how many input pairs are
needed to produce a chain of some fixed length with (almost) unity
probability of success?'' are asymptotically equivalent.

\item For {\it two-dimensional structures}, we prove that
        one can 
        build up cluster states with the optimal,
        quadratic use in
        resources, even when resorting to probabilistic gates:
        for any success probability $p_{\text s}\in(0,1]$
        of the physical primitive quantum gate, one can
        prepare a $n\times n$ cluster state consuming $O(n^2)$
				EPR pairs.
        Previously known schemes have operated with
        a more costly scaling.
        This is possible in a way that the overall 
        success probability
        $P_{\text s}(n) \rightarrow 1$ as $n\rightarrow \infty$. That is,
        even for quantum gates operating with a very small
        probability of success $p_{\text s}$, one can
        asymptotically {\it deterministically} build up
        two-dimensional cluster states using {\it quadratically
        scaling resources}.

\item For this preparation, we observe an intriguing
        {\it percolation effect} when preparing cluster states
        using probabilistic gates: 
        from a certain {\it threshold probability}
        \begin{equation*}
                p_{\text s}>p_{\text{th}}
        \end{equation*}
        on,
        {\it almost all} preparations of a $n\times n$ cluster
        will succeed, for large $n$.
        In turn, for
                $p_{\text s}<p_{\text{th}}$,
        {\it almost no} preparation will succeed
        asymptotically.

\item 	Also, cluster structures can be used for
				{\it loss tolerant} or {\it fully fault tolerant} quantum
				computing using linear optics. The required resources for the
				letter are tremendous, so the ideas presented here should give
				rise to a very significant reduction in the number of EPR
				pairs required.
\end{itemize}
In deriving the bounds, we assumed dealing with a linear optical
scheme 
\begin{itemize}
	\item   based on the {\it computational model of one-way computing} on
					cluster states in dual-rail encoding.
	\item using {\it EPR pairs from sources} as resource to build
					up cluster states, and allowing for any number of
					additional {\it vacuum modes} that could assist the
					quantum gates,        
	\item 
			such that one {\it sequentially} builds up the cluster
			state from elementary fusion {\it quantum gates}.       
\end{itemize}
Sequential means 
that we do not consider the possible multi-port devices 
-- as, e.\,g., in Ref.\ \cite{Port} --  involving a large number of
systems at a time (where the meaning of the asymptotic
scaling of resources is not necessarily well-defined).
In this sense, we identify the final limit of performance 
of such a linear optical architecture for quantum computing.

Structurally, we first discuss the physical setting.
After introducing a few concepts necessary for what
follows, we discuss on a more phenomenological level
the impact of the classical strategy on the resource
consumption \cite{Trees,Trees2}.
The longest part of the paper
is then concerned with the rigorous formal arguments.
Finally, we summarize what has been achieved, and
present possible scopes for further work in this direction.

\section{Preparing linear
cluster states with probabilistic quantum
gates}

\begin{figure}
  \includegraphics{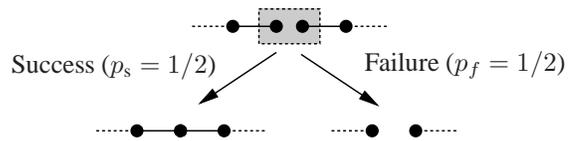}
  \caption{ Action of a fusion gate on the end qubits of two 
  linear cluster states. \label{fig:fusion1} }
\end{figure}

\subsection{Cluster states and fusion gates}

A {\it linear cluster state} \cite{Oneway}
is an instance of a
{\it graph state} \cite{GS,LongGS}
of a simple graph corresponding to a
line segment. Any such {\it graph state} is associated
with an undirected
graph, so with $n$ {\it vertices}  and a set $E$ of
{\it edges}, so of pairs $(a,b)$ of connected
vertices. Graph states can be defined as those states whose state
vector is of the form
\begin{equation*}
        |G\rangle = \prod_{(a,b)\in E}
        U^{(a,b)} \left((|0\rangle+ |1\rangle)/2^{1/2}\right)^{\otimes n}
\end{equation*}
where $
        U^{(a,b)} := |0\rangle\langle 0|^{(a)}
        \otimes \id^{(b)} +
        |1\rangle\langle 1|^{(a)} \otimes 
        \sigma_z^{(b)}$,
$\sigma_z$ denoting
the familiar Pauli operator. In this basis,
a linear cluster state
vector of some length $l$
is hence just a sum of all binary words on
$n$ qubits with appropriate phases. An {\it EPR pair}
is consequently
conceived as a linear cluster state with a single edge,
$l=1$ \cite{Chris}. 
A {\it two-dimensional cluster state} is the graph state
corresponding to a two-dimensional cubic lattice.
Only the describing graphs will be relevant in the sections to come;
the quantum nature of graph states does not enter our 
considerations.

As stated before, we call a quantum mechanical gate a (type-I)
\emph{fusion gate} \cite{terry} if it can ``fuse together'' two linear
cluster state ``chains'' with $l_1$ and $l_2$ edges respectively to
yield a single chain of $l_1+l_2$ edges (see Fig.~\ref{fig:fusion1}).
The process is supposed to succeed with some probability $p_\text{s}$.
In case of failure both chains loose one edge each: $l_i \mapsto
l_i-1$. Unless stated otherwise, we will assume that $p_\text{s}=1/2$,
in accordance with the results of the next section.

This kind of quantum gate is, yet, insufficient to build up
two-dimensional cluster states. For this to be possible, another kind
of fusion gate is required: \emph{type-II fusion}~\cite{terry} to be
discussed in Section \ref{2d}.

\subsection{Linear optical fusion gates}
\label{sc:FusionGates}

We use the usual convention for encoding a
qubit into photons: In the so-called
\emph{dual-rail} encoding the basis
vectors of the computational Hilbert space
are represented by
\begin{eqnarray*}
        |0\rangle &:=&a^\dagger_0|\text{vac}\rangle\\
        |1\rangle &:=&a^\dagger_1|\text{vac}\rangle,
\end{eqnarray*}
where $a^\dagger_{0,1}$
denote the creation operators in two orthogonal modes,
and $|\text{vac}\rangle$ is the state vector of the
vacuum.
The canonical choice are two modes that only differ in the
polarization degree
of freedom, e.\,g.
horizontal and vertical with respect to some reference,
giving rise to the notation
$|H\rangle :=|0\rangle$ and $|V\rangle :=|1\rangle$.

Type-I fusion gates were introduced in Ref.\ \cite{terry}, where it
was realized that the \emph{parity check gate}~\cite{PC} has exactly
the desired effect. The gate's probability of success is
$p_\text{s}=1/2$ and the following theorem states that this cannot be
increased in the setting of dual rail coded linear optical quantum
computation.

\begin{theorem}[Maximum probability of success of fusion]\label{Fuse}
The optimal probability of success $p_{\text s}$
of a type-I
fusion quantum gate is $p_{\text s}=1/2$. More specifically,
the maximal $p=p_1+p_2$ such that
\begin{eqnarray*}
        A_1 & =& p_1^{1/2}( |H\rangle\langle H,H| - |V\rangle\langle V,V| )/\sqrt{2},\\
        A_2 & =& p_2^{1/2} ( |H\rangle\langle H,H| + |V\rangle\langle V,V| )/\sqrt{2}
\end{eqnarray*}
are two Kraus operators of a channel that can be realized with
 making use of (i) any number of
 auxiliary modes prepared in the vacuum, (ii)
 linear optical networks acting on all modes,
 and (iii) photon counting detectors is given by
 $p=p_{\text s}:=1/2$.
\end{theorem}

\begin{figure}
  \includegraphics{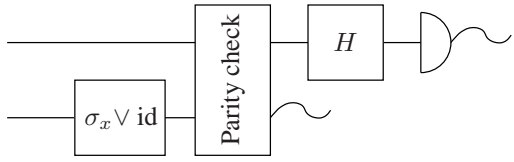}
  \caption{Diagram representing how 
  parity check gate can be employed to realize 
  a Bell state discriminating device.
    \label{fig:parity_bell} }
\end{figure}

\begin{proof}
   Given the setup in Fig.~\ref{fig:parity_bell}, we notice a parity check
  described by these Kraus operators can be used to realize a
  measurement, distinguishing with certainty
  two from four binary Bell states:
  The following Hadamard gate and measurement in the computational
  basis give rise to the Kraus operators
  \begin{equation*}
          B_{\pm}=\langle\pm|=2^{-1/2}(\langle H|\pm\langle V|).
  \end{equation*}
  On input of the symmetric Bell states with
  state vectors,
  $|\phi^{\pm}\rangle=2^{-1/2}(|H,H\rangle\pm|V,V\rangle)$,
  the measurement results $(A_1,B_-)$ and
  $(A_2,B_+)$ indicate a $|\phi^+\rangle$ and
  $(A_1,B_+)$ and $(A_2,B_-)$ a $|\phi^-\rangle$, respectively.
  These two states can be identified
  with certainty. The anti-symmetric
  Bell states with state vectors
  $|\psi^{\pm}\rangle=2^{-1/2}(|H,V\rangle\pm|V,H\rangle)$,
  will in turn result in a failure outcome.

  Applying a bit-flip (a Pauli $\sigma_x$)
  on the second input qubit (therefore implementing the
  map
   $|\phi^{\pm}\rangle \mapsto |\psi^{\pm}\rangle$,
   $ |\psi^{\pm}\rangle\mapsto |\phi^{\pm}\rangle$) at random,
  a discrimination between the
  four Bell states with uniform
  \emph{a priori} probabilities is possible,
  succeeding in 50\% of all cases.
  Following Ref.\ \cite{CL01} this is already the
  optimal success probability when only allowing for
  (i) auxiliary vacuum modes, (ii) networks of
  beam splitter and phase shifts and (iii)
  photon number resolving detectors. Thus,
  a more reliable parity check is not possible within
  the presented framework.
\end{proof}

In turn, it is straightforward to see that  a failure necessarily
leads to a loss of one edge each.  Note that one could in principle use
additional single-photons from sources or EPR pairs to attempt to
increase the success probability $p_{\text s}$ of the individual gate.
These additional resources would yet have to be included in the
resource count. Such a generalized scenario will not be considered.

\section{Concepts: Configurations and strategies}
\label{sc:Concepts}

The current section will set up a rigorous framework for the
description and assessment of control strategies. All considerations
concern the case of one-dimensional cluster states; the two
dimensional case will be deferred to Section \ref{2d}. Note that,
having described the action of the elementary gate on the level of
graphs, we may abstract from the quantum nature of the involved cluster
states altogether.

\subsection{Configurations}

A \emph{configuration} (in the \emph{identity picture}) $I$ is a list
of numbers $I_k, k\in\mathbbm{N}$.  We think of $I_k$ as specifying
the length of the $k$-th chain that is available to the experimenter
at some instance of time.  For most of the statements to come a more
coarse-grained point of view is sufficient: in general we do not have
to distinguish different chains of equal length. It is hence expedient
to introduce the \emph{anonymous representation} of a configuration
$C$ as a list of numbers $C_i, i\in \mathbbm{N}$ with $C_i$ specifying
the numbers of chains of length $i$. We will always use the latter
description unless stated otherwise.  Trailing zeroes will be
suppressed, i.\,e. we abbreviate $C=1,2,0,\dots$ as $C=(1,2)$.  Define
the \emph{total number of edges} (total length) to be $L(C)=\sum_i
i\,C_i$.  The space of all configurations  is denoted by
$\mathcal{C}$. By $\mathcal C^{(N)}$ we mean the set of configurations
$C$ having a total length less or equal to $N$.  Lastly, let $e_i$ be the
configuration consisting of exactly one chain of length $i$. This
definition allows us to expand configurations as $C=\sum_{i=1}^\infty
C_i\,e_i$.

\subsection{Elementary rule}

Let us re-formulate the action of the fusion gate in this language.
An attempted fusion of two chains of length $k$ and $l$ gives rise to
a map $C=\sum_{i=1}^\infty C_i e_i \mapsto C' = \sum_{i=1}^\infty C_i'
e_i$ with 
\begin{equation*}
  C' = C -e_k-e_l+e_{k+l}
\end{equation*}	
in case of success with probability $p_{\text s}=1/2$
(leading to a single chain of length $l+k$)
and
\begin{eqnarray*}
  C' = C -e_k+e_{k-1}-e_l+e_{l-1}
\end{eqnarray*}
in case of failure, meaning that one edge each is lost
for the chains of length $k$ and $l$. All other 
elements of $C$ are
left unchanged.

\subsection{Strategies}

A \emph{strategy} (in the anonymous picture) defines what \emph{action} to take when faced with a
specific configuration. Actions can be either ``try to fuse a chain of
length $k$ with one of length $l$'' or ``do nothing''. Formally, we will
represent these choices by the tuple $\langle k, l \rangle$ and
the symbol $\stale$, respectively. 
It is easy to see that, in trying to build up a single long chain, it
never pays off not to use all available resources. We hence require a
strategy to choose a non-trivial action as long as there is more than
one chain in the configuration. Formally, a strategy is said to be
\emph{valid} if it fulfills
\begin{enumerate}
	\item (No null fusions): 
	$S(C) = \langle k, l \rangle \Rightarrow C_l,C_k\neq 0$

	\item (No premature stops):
	$S(C) = \stale \Leftrightarrow$ $C$ contains at most one chain.
\end{enumerate}
We will implicitly assume that all strategies that appear are valid.
Strategies in the identity picture are defined completely analogously.

An event $E$ is a string of elements of $\{S, F\}$, denoting success
and failure, respectively. The $i$-th component of $E$ is denoted by
$E_i$ and its length by $|E|$. Now fix an initial configuration
$C_\emptyset$ and some strategy $S$. We write $C_E$ for the
configuration which will be created by $S$ out of $C_\emptyset$ in the
event $E$. Here, as in several definitions to come, the strategy $S$
is not explicitly mentioned in the notation.  It is easy to see that
any strategy acting on some initial configuration will, in any event,
terminate after a finite number of steps $n_{\text T}(C)$.

Recall that the outcome of each action is probabilistic and \emph{a
priori} we do not know which $C_E$ with $|E|=n$ will have been
obtained in the $n$-th step. It is therefore natural to introduce a
probability distribution on $\mathcal{C}$, by setting
\begin{equation*} 
	p_n(C):=2^{-n}\,|\{ E: |E|=n, C_E=C \}|.
\end{equation*} 
In words: $p_c(C)$ equals $2^{-n}$ times the number of events that
lead to $C$ being created.
The fact that $S$ terminates after a finite number of steps translates
to $p_{n_{\text T}+k}=p_{n_{\text T}}$ for all positive integers $k$.
Expectation values of functions $f$ on $\mathcal{C}$ now can be written as
\begin{equation*}
  \langle f\rangle_{p_n} : = \sum\limits_{C} p_n(C)f(C).
\end{equation*}
The \emph{expected total length} is
\begin{equation*}
  \langle L\rangle_{p_n} := \sum\limits_{C,i}p_n(C)i\,C_i.
\end{equation*}
In particular, the \emph{expected final length} is given by
$\tilde{Q}(C_\emptyset) := \langle L\rangle_{p_{n_{\text T}}}$. 
Of central importance will be the best possible expected final length
that can be achieved by means of any strategy:
\begin{equation*}
	Q(C_\emptyset) := 
	\sup_{S}\tilde{Q}_S(C_\emptyset).
\end{equation*}
This number will be called the \emph{quality} of $C_\emptyset$.  For
convenience we will use the abbreviations $\tilde Q(N):=\tilde Q(N e_1)$ and $Q(N):=Q(N e_1)$. \\

\section{Simple strategies}
\begin{figure}
  \center \includegraphics{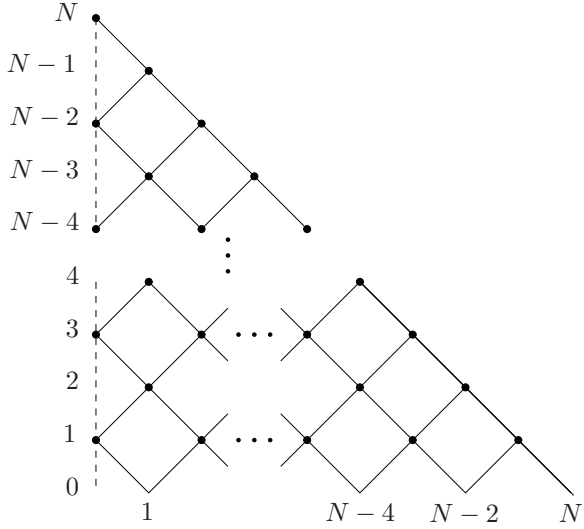}
	\caption{ The process of fusion of the largest can be represented as a tree
	similar to a random walk. Reflection occurs at the dashed line (the largest string is
	lost and replaced with an EPR pair. Time evolves from top to bottom,
	thus decreasing the number of EPR resources. The horizontal
	dimension represents the length of the largest string.
    \label{fig:random_walk} }
\end{figure}

A priori, a strategy does not allow for a more economic description
other than a 'look-up table', specifying what action to take when
faced with a given configuration. If one restricts attention to the
set of configurations $\mathcal{C}^{(N)}$ that can be reached starting
from $N$ EPR pairs, $|\mathcal{C}^{(N)}|$ values have to be fixed.

The cardinality $|\mathcal{C}^{(N)}|$, in turn, can be derived from
the accumulated number of integer partitions of $k\le N$.  The
asymptotic behavior \cite{sloane} can be identified to be
\begin{equation*}
  |\mathcal{C}^{(N)}| = \frac{1+O(N^{-1/6})}{({8\pi^2N})^{1/2}}\me^{\pi({2N}/{3})^{1/2}} ,
\end{equation*}
which is exponential in the number $N$ of initially available EPR pairs~\cite{Kieling05a}.

However, there are of course strategies which do allow for a simpler
description in terms of basic general rules that apply similarly to
all possibly configurations. It might be surmised that
close-to-optimal strategies can be found among them. Also, these simple
strategies are potentially accessible to analytical and numerical
treatment.  Subsequently, we will discuss three such reasonable
strategies, referred to as {\sc Greed}, {\sc
Modesty}, and {\sc Static}.

\subsection{\sc Greed}
\begin{figure}
  \includegraphics{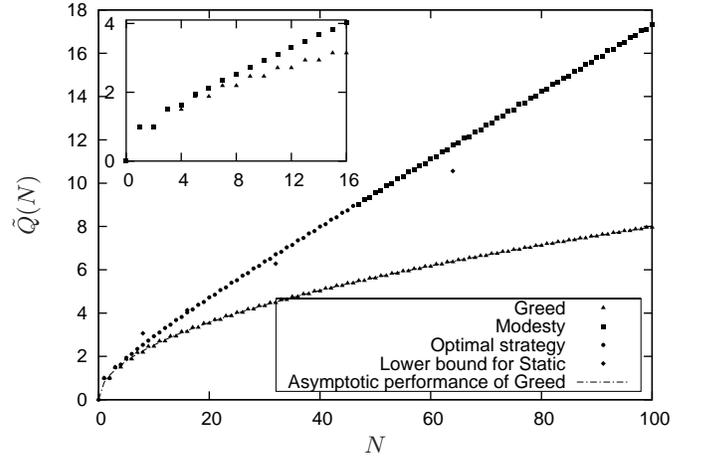}
  \caption{ Expected length for the globally optimal strategy,
    for {\sc Modesty} (in this
    plot indistinguishable from the former),
    for {\sc Greed}, its asymptotic performance,
    and the lower bound for {\sc Static}, as
    functions of even number $N$ of initial EPR pairs. The inset shows {\sc Greed} and {\sc Modesty}
    for small $N$, revealing the parity-induced step-like behavior.
    \label{fig:strategies} 
    }
\end{figure}

This is one of the most intuitive strategies. It can be described as
follows:  ``Given any configuration, try to fuse the largest two
available chains''. This is nothing but
\begin{equation*}
  S_G(C) = \left\{\begin{array}{ll}
           \emptyset & \text{if $\sum_iC_i \le 1$} \\
           \langle k,l\rangle & \begin{array}{lcl} k&=&\max\{i:C_i > 0\} \\ l&=&\max\{i:C_i-\delta_{i,k} > 0\}\end{array}
         \end{array} \right. .
\end{equation*}
Alternatively, one may think of {\sc Greed} as fusing the first two
chains after sorting the configuration in descending order. 
The rationale behind choosing this strategy is the following: fusing
is a probabilistic process which destroys entanglement on average.
Hence it should be advantageous to quickly build up as long a chain as
possible. Clearly, the strategy's name stems from its pursuit of
short-term success. From a theoretical point of view, {\sc Greed} is
interesting, as its asymptotic performance can easily be assessed
(see Fig.~\ref{fig:strategies}):

\begin{lemma}[Asymptotic performance of 
{\sc Greed}]\label{lemma:greed}
 The expected length of the final chain after 
 applying {\sc Greed} to $N$ EPR pairs scales 
  asymptotically as 
  \begin{equation*}
    \tilde Q_G(N) = ({2N}/{\pi})^{1/2} + o(1).
  \end{equation*}
\end{lemma}

\begin{proof} 
	It is easy to see that an application of
	{\sc Greed} to $C_\emptyset = Ne_1$ only generates configurations in
	$\bigl\{me_1+e_l, m=0,\ldots,N; l=0,\ldots,N; l+m\le N\bigr\}$. This
	set is parametrized by $m$ (the number of EPR resources) and $l$
	giving rise to the notation $C=(l,m)$. By definition of $S_G$,
	whenever $l\ge1$, the next fusion attempt is made on this longer
	chain and one of the other EPR pairs. As for the case $l=0$ we
	identify $(0,m)$ with $(1,m-1)$ (when encountering $(0,m)$ we
	distinguish one of the $m$ pairs).  Therefore, in this slightly
	modified notation we have with $C_E = (l,m), l>0$ in case of success
	$C_{ES}=(l+1,m-1)$ and in case of failure $C_{EF}=(l-1,m-1)$,
	respectively.

  The tree in Fig.~\ref{fig:random_walk} can be obtained by reflecting the negative half of a standard random
  walk tree at $l=0$ and identifying the vertices with same $m$ but
   opposite $l$. One can readily read off the
  expectation value of final chain's length. The form is especially simple
  in the balanced case ($p_{\text s}=1/2$),
  \begin{equation*}
    \tilde Q_G(N) = 
    2 \sum\limits_{k=0}^{\left\lfloor(N-1)/2\right\rfloor}p_{\text s}^k(1-p_{\text s})^{N-k}{N\choose k}\left(N-2k\right).
  \end{equation*}
  The probabilities are twice the probabilities of the standard random
  walk tree, and the length-$0$ term has been omitted.

  Using an estimate using a Gaussian distribution 
  we easily find the asymptotic behavior for large $N$
  (setting $\mu=p N$ and $\sigma^2=p_{\text s} (1-p_{\text s}) N$ with $p_{\text s}=1/2$),
  \begin{eqnarray*}
    \tilde Q_G(N)
      &=& \left({\frac{8}{N\pi}}\right)^{1/2}\int\limits_0^{\infty}
      2x\exp\left(-\frac{2x^2}{N}\right)\md x + r(N) \nonumber \\
      &=& \left(\frac{2N}{\pi}\right)^{1/2}
      \Gamma(1) + r(N)
  \end{eqnarray*}
  with approximation error $r(N)=o(1)$.
\end{proof}

\begin{figure}
  \includegraphics{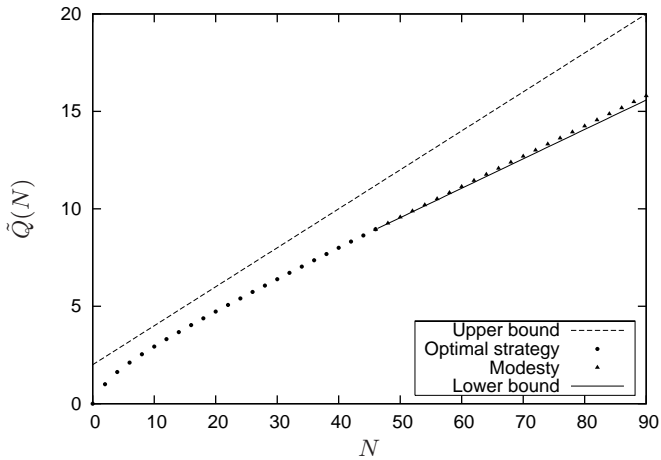}
  \caption{ Expected length for {\sc Modesty}, the optimal strategy (where known)
    a lower bound to the quality as in Theorem~\ref{tmLowerBound}, but with $N_0=46$ (for better visualization),
    and the upper bound attained with the razor model as functions of the number of initial EPR pairs $N$. \label{fig:bounds} }
\end{figure}

The behavior of {\sc Greed} changes qualitatively upon variation of
$p_{\text s}$: For $p_{\text s}>1/2$, $\tilde Q_G(N)$ shows linear
asymptotics in $N$, while in case of $p_{\text s}<1/2$ the quality
$\tilde Q_G(N)$ is not even unbounded as a function of $N$.

There is a phenomenon present in the performance of many strategies,
which can be understood particularly easily when considering {\sc
Greed}: $\tilde Q$ displays a ``smooth'' behavior when regarded as a function
on either only \emph{even} or only \emph{odd} values of $N$. However, the
respective graphs appear to be slightly displaced with respect to each
other. For simplicity, we will in general restrict our attention to even
values and explore the reasons for this behavior in the following
lemma.

\begin{lemma}[Parity and $\tilde Q_G$]
\label{lm:parity}
	Let $N$ be even. Then $\tilde Q_G(N)=\tilde Q_G(N+1)$.
\end{lemma}

\begin{proof}
	Let $C_\emptyset=N e_1, C'_\emptyset= (N+1) e_1$, for $N$ even.  Now
	let $E$ be such that $S(C_E)=\stale$ but
	$S(C_{E_{1,\ldots,|E|-1}})\ne\stale$.  As {\sc Greed} does not touch
	the $i$-th chain before the $i$-th step, it holds that
	$C'_E=C_E+e_1$. Further, since type-I fusion preserves the parity of
	the total number of edges, $C'_E\neq 0$. Hence $C'_E$ is of the form
	$C'_E=e_1+e_k$ and one computes:
	\begin{eqnarray*}
		\tilde{Q}_G(C'_E)
		&=& 1/2 (k+1) + 1/2 (k-1) = k = \tilde Q_G(C_E).
	\end{eqnarray*}
	From here, the assertion is easily established by re-writing $\tilde
	Q_G(C'_\emptyset)$ as a  suitable average over terms of the form
	$Q_G(C'_E)$, where $E$ fulfills the assumptions made above.
\end{proof}

As a corollary to the above proof, note that fusing an EPR pair to
another chain does not, on average, increase its length. Hence the
fact that $\tilde Q_G(N)$ grows at all as a function of $N$ is solely
due to the asymmetric situation at length zero.

Lemma \ref{lm:parity} explains the steps apparent in
Fig.~\ref{fig:strategies}. Such steps are present also in the
performance of {\sc Modesty}, to be discussed now, and several other
strategies -- albeit not in such a distinct manner.

\subsection{\sc Modesty}

There is a very natural alternative to the previously studied
strategy.  Instead of trying to fuse always the largest existing
linear cluster states in a configuration, one could try the opposite:
``Given any configuration, try to fuse the smallest two available
chains''.  In contrast to {\sc Greed} this strategy intends to build up
chains of intermediate length, making use of the whole EPR reservoir
before trying to generate larger chains.  Even though no long chains
will be available at early stages, the strategy might nevertheless
perform reasonably.  Quite naturally, this strategy we will call {\sc
Modesty}.

Formally, this amounts to replacing $\max$ by $\min$, i.\,e. replacing
descending order by ascending order:
\begin{equation*}
  S_M(C) = \left\{\begin{array}{ll}
           \emptyset & \text{if $\sum_iC_i \le 1$} \\
           \langle k,l\rangle & \begin{array}{lcl} k&=&\min\{i:C_i > 0\} \\ l&=&\min\{i:C_i-\delta_{i,k} > 0\}\end{array}
         \end{array} \right. .
\end{equation*}
Maybe surprisingly, {\sc Modesty} will not only turn out to give
better results than {\sc Greed}, but is actually close to being
globally optimal, as can be seen in Figures~\ref{fig:strategies}
and~\ref{fig:bounds}. See Section \ref{sc:Data} for a closer
discussion.

\subsection{\sc Static}

Another strategy of particular interest is called {\sc Static}, $S_S$.
To describe its action, we need to define the notion of an
\emph{insistent strategy}. The term is only meaningful in the identity
picture, which we will employ for the course of this section. Now, a
strategy is called insistent if, whenever it decides to fuse two
specific chains, it will keep on trying to glue these two together
until either successful or at least one of the chains is completely
destroyed.  Formally:
\begin{equation*} 
  S(C_E)=\langle k,l\rangle\wedge (C_{EF})_k(C_{EF})_l \ne 0
  \Rightarrow S(C_{EF}) = S(C_E)
\end{equation*}

{\sc Static} acts by insistingly fuse the first chain to the second
one; the third to the fourth and so on.  After this first level, the
resulting chains will be renumbered in the way that the outcome of the
$k$-th pair is now the $k$-th chain.  At this point, {\sc Static}
starts over again, using the configuration just obtained as the new
input. This procedure is iterated until at most one chain of nonzero
length has survived.

The proceeding of $S_S$ is somehow related to {\sc Modesty} and {\sc
Greed}, just without sorting the chains between fusion attempts.  This
results in much less requirements on the routing of the photons
actually carrying the cluster states. From an experimentalist's point
of view, {\sc Static} is a meaningful choice as it only requires a
minimal amount of classical feed-forward that is only present at the
level of fusion gates, not on the level of routing the chains.  It
performs, however, asymptotically already better than {\sc Greed} (see
Fig.~\ref{fig:strategies}).

It turns out that {\sc Static} performes rather poorly when acting on
a configuration consisting only of EPR pairs. To cure this deficit, we
will proceed in two stages. Firstly, the input is partitioned into
blocks of eight EPR pairs each. Then {\sc Modesty} is used to
transform each block into a single chain. The results of this first
stage are subsequently used as the input to {\sc Static} proper, as
described before.  Slightly overloading the term, we will call this
combined strategy {\sc Static} as well. Note that, even when
understood in this wider sense, {\sc Static} still reduces the need
for physically re-routing chains: the blocks can be chosen to consist
of neighboring qubits and no fusion processes between chains of
different blocks are necessary during the first stage.  The following
theorem bounds {\sc Static}'s performance.  For technical reasons, it
is stated only for suitable $N$.

\begin{theorem}[Linear performance of {\sc Static}] \label{theorem:static}
	For any $m\in \mathbbm{N}$, given $N=2^{3+m}$ EPR pairs, {\sc
	Static} will produce a single chain of expected length
  \begin{equation*}
    \tilde Q(N) \ge (137/1024) N + 2 .
  \end{equation*}
\end{theorem}

The proof of the above theorem utilizes the following lemma which
quantifies the quality one expects when combining several configurations.

\begin{lemma}[Combined configurations]
\label{lmAddingConfigs}
  The following holds.
  \begin{enumerate}
    \item
      Let $C$ be a configuration consisting of single chains of respective
      length $l_1, l_2$. Then~\cite{twoChainsGeneral}
      \begin{equation}\label{twoChains}
        Q(C)=l_1+l_2 - 2 + 2^{1-\min(l_1,l_2)} \ge l_1+l_2-2.
      \end{equation}

    \item
      Let $C_{(1)},\dots, C_{(k)}$ be configurations. Let $S$ be a strategy
      that acts on $\sum_iC_{(i)}$ by first acting with $S'$ on each of the
      $C_{(i)}$ and then acting insistently on the resulting chains. Then~\cite{kChainsGeneral},
      \begin{equation*}
        \tilde Q_{S}\left(\sum_i C_{(i)}\right)\geq \sum_i \tilde Q_{S'}(C_{(i)}) - 2(k-1) .
      \end{equation*}

    \item 
		When substituting all occurences of $\tilde Q$ by $Q$, the above
		estimate remains valid.
  \end{enumerate}
\end{lemma}

\begin{proof}
	Firstly, any strategy will try to fuse the only two chains in the
	configuration together until it either succeeds or the shorter one
	of the two is destroyed (after $\min(l_1,l_2)$ unsuccessful
	attempts). In other word: in case of these special configurations
	any strategy is insistent. By Lemma~\ref{lm:Attempts}:
  \begin{eqnarray*}
    Q(l_1,l_2) 
      &=& l_1+l_2 - \langle T \rangle  \nonumber\\
      &=& l_1+l_2 - \sum_{i=0}^{\min(l_1,l_2)-1} 2^{-i} \nonumber\\
      &=& l_1+l_2 - 2 + 2^{1-\min(l_1,l_2)}.
  \end{eqnarray*}

	For the second part, we run $S'$ on each $C_{(i)}$, resulting in $k$
	single chain configurations $C'_{(i)}=e_{l_i}$ with probability
	distributions $p_i$ on $\mathcal{C}$ obeying $\tilde
	Q_{S'}(C'_{(i)}) = \langle l_i\rangle_{p_i}$.  The joint
	distribution on $\mathcal C^k$ is given by $p=\prod_i p_i$. Now we
	fuse the chains together. If $C'_{(i)}$ and $C'_{(j)}$ are such that
	$p_i(C'_{(i)})p_j(C_{(j)})\ne0$, we unite them into one
	configuration $C:=C_{(i)} + C_{(j)}$. Clearly, $C$ contains at most
	two chains which we fuse together as described in the first part of
	the Lemma. As Eq. (\ref{twoChains}) is linear in the respective
	lengths of the chains in $C$, the distribution $p'=p_i p_j$ fulfills
	on the one hand
  \begin{equation*}
    \langle \tilde Q \rangle_{p'} = 
		\langle L\rangle_{p'} \ge \langle L\rangle_{p_i}+\langle L\rangle_{p_j}-2
  \end{equation*}
  and on the other hand
  \begin{equation*}
    \tilde Q( (\langle l_i\rangle_{p_i}, \langle l_j\rangle_{p_j}) ) \ge 
		\langle L\rangle_{p_i}+\langle L\rangle_{p_j}-2.
  \end{equation*}
	for any insistent strategy. Because these two quantities are
	bounded by the same value we will use this bound and replace
	averages over $\tilde Q$ with $\tilde Q$ of configurations of
	average lengths.

	We now iterate this scheme to obtain a single chain. A moment of
	thought reveals that -- as a result of our neglecting the
	$2^{1-\min(l_1,l_2)}$-term -- the order in which chains are
	fused together does not enter the estimate for $\tilde Q_S$. The
	claim follows.

  As for the third point: It follows by setting $S'$ to the optimal strategy.
\end{proof}

\begin{proof}\emph{(of Theorem~\ref{theorem:static})}
	Consider a configuration consisting of $n=2^m$ chains of length $x$
	each. Using Lemma~\ref{lmAddingConfigs} one sees that the second
	stage of {\sc Static} will convert it into a single chain of
	expected length $\tilde Q(2^m e_x) \geq (x-2)n+2$.
	
	According to Section~\ref{sc:ComputerResults}, {\sc Modesty} fulfils
	$\tilde Q_M(8)=Q(8)=649/256$. Applying 
	Lemma~\ref{lmAddingConfigs} again we find with $x=\tilde Q_M(2^3)$ and $N=2^{3+m}$
  \begin{eqnarray*}
		\tilde Q_S(N) &\ge& \frac{649/256-2}{8}N+2 = \frac{137}{2048}N+2
		\\&\approx& 6.69\ 10^{-2}N+2.
  \end{eqnarray*}
\end{proof}

In case of $p_{\text s}\ne1/2$,
\begin{equation*} 
	\tilde Q'_S(ne_x) \ge 
	n(x-l_{\text{initial}})+l_{\text{initial}} 
\end{equation*}
can be obtained in the same way, where
$l_{\text{initial}}=2(1-p_{\text s})/p_{\text s}$ (similar to $n_c$
in~\cite{DR}).  Initial chains of length $\ge l_{\text{initial}}$ can
be produced by employing for example {\sc Greed}, but disregarding the
outcome in case of a fusion failure and aborting the process when
$2(1-p_{\text s})/p_{\text s}$ is reached. Although large chains are
produced with only a small overall success probability, this does not
effect the linear asymptotics as this process only depends on
$p_{\text s}$, rather than $N$.

\section{Computer-assisted results}
\label{sc:ComputerResults}

\subsection{Algorithm for finding the optimal strategy}

Before passing from the concrete examples considered so far to the
more abstract results of the next sections, it would be instructive to
explicitly construct an optimal strategy for small $N$. Is that a
feasible task for a desktop computer? Naively, one might expect it
not to be. Since the number of strategies grows super-exponentially 
as a function of the total number of edges $N$ of the initial
configuration, a direct comparison of the strategies' performances is
quickly out of reach. Fortunately, a somewhat smarter, recursive
algorithm can be derived which will be described in the following
paragraph.

The \emph{number of vertices} in a configuration is given by
$V(C):=\sum_i C_i (n_i+1)$. An attempted fusion will \emph{decrease}
$V(C)$ regardless of whether it succeeds or not.  Now fix a $V_0$ and
assume that we know the value of $Q$ for all configurations comprised
of up to $V_0$ vertices. Let $C$ be such that $V(C)=V_0+1$. It is
immediate that 
\begin{equation*}
 Q(C)=\text{max}_{i,j}\,
(Q(\text{S}_{i,j}\,C)+Q(\text{F}_{i,j}\,C))/2,  
\end{equation*}
where
$\text{S}_{i,j}\,C$ denotes the configuration resulting from
successfully fusing chains of lengths $l_i$ and $l_j$.
$\text{F}_{i,j}C$ is defined likewise.  As the r.\,.h.\,s.\ involves only
the quality of configurations possessing less than or equal to $V_0$ vertices, we
know its value by assumption and we can hence perform the maximization
in $O(c^2)$ steps. One thus obtains the quality of $C$ and the pair of
chains that need to be fused by an optimal strategy.

The algorithm now works by building a \emph{lookup table} containing
the value of $Q$ for \emph{all} configurations up to a specific
$V_{\text{max}}$. It starts assessing the set of configurations with
$V(C)=1$ and works its way up, making at each step use of the
previously found values. One needs to supply an anchor for the
recursion by setting $Q(e_i)=i$. Clearly, the memory consumption is
proportional to $|{\cal C}^{(N)}|$, which is exponential in $N$ and will limit
the practical applicability of the algorithm before time issues do.

We have implemented this algorithm using the computer algebra system
{\it Mathematica} and employed it to derive in closed form an optimal
strategy for all configurations in ${\cal C}^{(46)}$, the quality of
which is shown in Figures~\ref{fig:strategies} and~\ref{fig:bounds}.
A desktop computer is capable of performing the derivation in a few
hours \cite{table}.  

From the discussion above, it is clear that the leading term in the
\emph{computational complexity} of the algorithm is given by
$|C^{(N)}|$: every configuration needs to be looked at at least once.
A straight-forward analysis reveals a poly-log correction; the
described program terminates after $O\big(|C^{(N)}|\,(\log
|C^{(N)}|)^5\big)$ steps. 

\subsection{Data, intuitive interpretation, and competing tendencies}
\label{sc:Data}

Starting with $C_\emptyset=N e_1$, {\sc Modesty} turns out to be the optimal
strategy for $N\leq 10$. For configurations containing more edges,
slight deviations from {\sc Modesty} can be advantageous. The
difference relative to $Q(N)$ is smaller than $1.1\times 10^{-3}$ for
$N\leq 46$. More generally, two heuristic rules seem to hold:
\begin{enumerate}
	\item
	It is favorable to fuse small chains (this is the dominant rule).

	\item
	It is favorable to create chains of equal length.
\end{enumerate}
Is there an intuitive model which can explain these findings? Several
steps are required to find one. Firstly, note that every fusion attempt
entails a $1/2$ probability of failure, in which case two edges are
destroyed. So ``on average'' the total length $L(C)$ decreases by one in
each step and it is natural to assume that \emph{the quality $Q(C)$
equals $L(C)$ minus the expected number of fusion attempts} a specific
strategy will employ acting on $C$. Hence a good strategy aims to
\emph{reach a single-chain configuration as quickly as possible}, so
as to reduce the expected number of fusions (this reasoning will be
made precise in Section~\ref{Upper}). 
Now, if there are $k$ chains present
in $C$, then a priori $k-1$ successful fusions are needed before a
strategy can terminate. If, however, in the course of the process one
chain is completely destroyed, then $k-2$ successes would already be
sufficient. Therefore -- paradoxically -- within the given framework
\emph{it pays off to destroy chains}. Since shorter chains are more
likely to become completely consumed due to failures, they should be
subject to fusion attempts whenever possible. This explains the first
rule. 

There is one single scenario in which \emph{two} chains can be
destroyed in a single step; that is when one selects two EPR pairs to
be fused together. Now consider the case where there are two chains of
equal length in a configuration. If we keep on trying to fuse these
two chains, then -- in the event of repeated failures -- we will
eventually be left with two EPR pairs, which are favorable to obtain
as argued before.  Hence the second rule.

We have thereby identified two \emph{competing tendencies} of the
optimal strategy.  Obtaining a quantitative understanding of their
interplay seems extremely difficult: deviating from {\sc Modesty} at
some point of time might open up the possibility of creating two
chains of equal lengths many steps down the line. We hence feel it is
sensible to conjecture that \emph{the globally optimal strategy allows
not even for a tractable closed description}. A \emph{proof} of its
optimality seems therefore beyond any reasonable effort. One is left
with the hope of obtaining appropriately tight analytical bounds --
and indeed, the sections to come pursue this programme with perhaps
surprising success.

\section{Lower bound}\label{Lower}
We will now turn to establishing rigorous upper and lower bounds
to $Q$, so the quality of the optimal strategy. These bounds, in turn,
give rise to bounds to the resource consumption any linear optical
scheme will  have to face. Lower bounds are in turn less technically
involved than upper bounds. In fact, rigorous lower bounds can
be based on known bounds for given strategies:
For not too-large configurations, the performance of various
strategies can be calculated explicitly on a computer (see Section
\ref{sc:ComputerResults}). Any such computation in turn gives a lower
bound to $Q$. The following theorem is based on a construction which
utilizes the computer results to build a strategy valid for inputs of
arbitrary size. This strategy is simple enough to allow for an
analytic analysis of its performance while at the same time being
sufficiently sophisticated to yield a very tight lower bound for the
quality, shown in Fig.~\ref{fig:bounds}. Notably, the resulting 
statement is {\it not} a numerical estimate
valid for small $N$, but a proven bound valid for all $N$:

\begin{theorem}[Lower bound for globally optimal strategy]
\label{tmLowerBound}
Starting with $N$ EPR pairs and using fusion gates, the
globally optimal strategy yields a cluster state of expected
length 
\begin{equation} \label{eqn:lowerbound}
  Q(N) \geq \tilde Q(N_0) + \alpha(N-N_0),
\end{equation}
for all $N\ge N_0$. The constants are 
\begin{eqnarray*}
	N_0&=&92, \qquad \tilde Q(N_0)=16.1069, \nonumber\\
	\alpha&=&(\tilde Q(N_0)-2)/N_0=0.153336.
\end{eqnarray*}
Rational expression are known and can be accessed at 
Ref.\ \cite{table}.
\end{theorem}

\begin{proof}
	Denote by $\tilde Q(N)$ the expected final length of some strategy
        acting on $N$ EPR pairs. Fix $N_0$ such that $\tilde Q(N)$ is known
        for all $N\leq 2N_0$ and $\tilde Q$
	satisfies for $N_0 \leq N \leq 2 N_0$
	\begin{equation} \label{eqn:propmod}
		(\tilde Q(N)-2)/{N} \geq (\tilde Q(N_0)-2)/N_0
	\end{equation}
	and that Eqn.~(\ref{eqn:lowerbound}) holds for all $N\leq 2 N_0$.

	Now assume we are given $N>2 N_0$ EPR pairs. Clearly, there are
	positive integers $k\geq 2$ and $M\leq N_0$ such that $N=k N_0 + M$.
	Set $n_i = N_0$ for $i=1,\dots,k-1$ and $n_k=N_0+M$.  The $n_i$
	fulfill $\sum_i n_i=N$ and $N_0 \leq n_i \leq 2 N_0$. We partition the
	input into blocks of length $n_i$ each and compute
	\begin{eqnarray*}
		Q\left(\sum_i n_i\right)
		&\geq& \sum_{i=1}^k Q(n_i) - 2(k-1) \nonumber\\
		&\geq& \sum_{i=1}^k \tilde Q(n_i) - 2(k-1)  \nonumber\\
		&=& \tilde Q(N_0) + \sum_{i=2}^k n_i \frac{\tilde Q(n_i) - 2}{n_i}  \nonumber\\
		&\geq& \tilde Q(N_0) + \sum_{i=2}^k n_i \frac{\tilde Q(N_0) - 2}{N_0}  \nonumber\\
		&=& \tilde Q(N_0) + \alpha \sum_{i=2}^k n_i \nonumber\\
		&=& \tilde Q(N_0) + \alpha (N-N_0),
	\end{eqnarray*}
	where we made use of Lemma~\ref{lmAddingConfigs} and the assumptions
	mentioned above.

	In the case of {\sc Modesty} the function $\tilde Q_M(N)$ can be
	explicitly computed for not too large values of $N$. Indeed, the
	results for all $N\leq 2 N_0=184$ can be found at \cite{table}.
	They obey the condition in Eqn.~(\ref{eqn:propmod}) and the statement
	follows with $\tilde Q_M(N_0)=16.1069$.
\end{proof}

\section{Upper bounds}\label{Upper}

While the performance of any strategy delivers a lower bound for the
optimal one, giving an upper bound is considerably harder. We will
tackle the problem by passing to a family of simplified models. For
every integer $R\geq 2$, the \emph{razor model with parameter $R$} is
defined by introducing the following new rule: after every fusion
step all chains will be cut down to a maximum length of $R$.
Obviously, the full problem may be recovered with $R\ge N$. Given the
complexity of the problem, it comes as a surprise that even for
parameters as small as $R=2$ the essential features of the full
setup seem to be retained by the simplification, in the sense that
understanding the razor model yields extraordinary good bounds for
$Q$.

\subsection{The razor model -- outline}
\begin{figure}
  \includegraphics{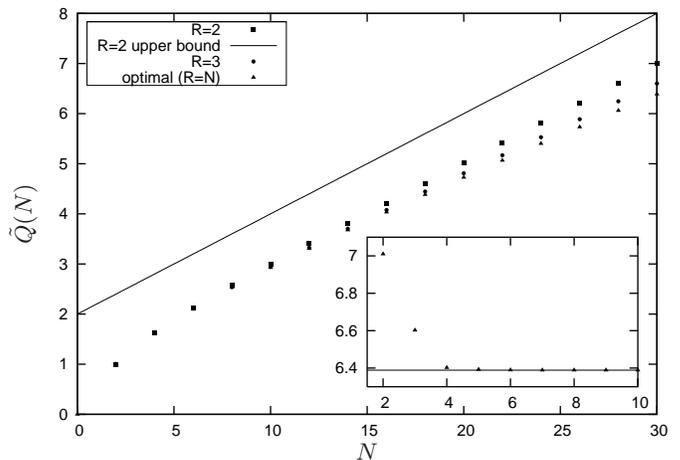}
  \caption{Performance of the optimal strategy in the razor model ($R=2$ and $R=3$), the full model ($R=N$) and the upper bound attained with the $R=2$ razor model.
    The inset shows the convergence of the upper bound to the quality (based on the razor model with parameter $R$) vs.
    the razor parameter $R=2,\ldots,10$ for $N=30$ together with the optimal value $Q(30)$.
    \label{fig:razor} }
\end{figure}
In the spirit of Section \ref{sc:Concepts}, a configuration in the
razor model is specified by a vector in $\mathbbm{N}^R$.  Thus, the
number of configurations with a maximum total number of $N$ edges is
certainly smaller than $N^R$, which is a polynomial in $N$. Adapting
the techniques presented in Section \ref{sc:ComputerResults}, we can
obtain the optimal strategy with polynomially scaling effort. \emph{We
have thus identified a family of simplified problems which, in the
limit of large $R$, tend to become exact, and where each instance is
solvable in polynomial time.}

How do the results of the razor model relate to the original problem?
Clearly, for small values of $R$, $Q_{\text{razor}}(C)$ will be a very
crude lower bound to $Q(C)$. However, as indicated in Section
\ref{sc:ComputerResults},
the quality of a configuration $C$ can be assessed in terms of the
optimal strategy's expected number of fusion attempts $\langle
T(C)\rangle$ when acting on $C$. It is intuitive to assume that
$\langle T \rangle \leq \langle T \rangle_{\text{razor}}$, as
the ``cutting process'' increases the probability of early termination.
We will thus employ the following argument: for
a given configuration $C$, derive a lower bound for $\langle T(C)
\rangle_\text{razor}$, which is in particular a lower bound for
$\langle T(C)\rangle$, which in turn gives rise to the upper bound
\begin{equation*}
	Q(C)\leq L(C) - \langle T(C) \rangle
\end{equation*}	 
for $Q$. 

The results of this ansatz are extremely satisfactory. Fig.~\ref{fig:razor}
shows the performance of the optimal quality for various $R$,
and the convergence when increasing $R$.

The intuitive explanation for the success of the model is the
observation that the chance that a chain of length $R$ is built up,
and eventually again disappears, is exponentially suppressed as a
function of $R$.  That is, the crucial observation is that the error
made by this radical modification is surprisingly small.  A rigorous
justification for this reasoning is supplied by the following two
propositions which will be proved in the next section.

\begin{lemma}[Quality and attemped fusions]
	\label{lm:Attempts}
	The expected final length $\langle L \rangle$ equals the initial
	number of edges $L(C_\emptyset)$ minus the expected number of attempted
	fusions $\langle T \rangle$.
\end{lemma}

\begin{theorem}[Bound to the full model from the razor model]
	\label{tm:Razor}
	Let $C_\emptyset\in\mathcal{C}$ be a configuration. The optimal strategy in
	the setting of the razor model will use fewer fusion attempts on
	average to reach a final configuration starting from $C_\emptyset$ than will
	the optimal strategy of the full setup.
\end{theorem}

\subsection{The razor model -- proofs}

For the present section, it will prove advantageous to introduce some
alternative points of view on the concepts used so far.  Recall that a
strategy is a function from \emph{configurations} to \emph{actions}.
However, once we have fixed some initial configuration $C_\emptyset$,
we can alternatively specify a strategy as a map from \emph{events} to
actions. Indeed, the configuration present after $n$ steps is
completely fixed by the knowledge of the initial configuration, the
past decisions of the strategy and the succession of failures and
successes. We will call the resulting mapping the \emph{decision
function} $D_{S,C_\emptyset}$ and will suppress the indices whenever no
danger of confusion can arise. In the same spirit, we are free to
conceive \emph{random variables} on $\mathcal C$ as real functions $f: \{S,F\}^n \to
\rr$.
Expectation values are then computed as
\begin{equation*}
	\langle f \rangle := \langle f \rangle (p_{n_{\text T}})
	\sum_{E,|E|=n_{\text T}} 2^{-|E|} f(E).
\end{equation*}
Quantities of the form $\langle f \rangle(C)$ for some configuration $C$
refer expectation values $\langle f \rangle$ given the initial
configuration $C_\emptyset=C$.

An interesting class of random variables can be written in the form 
\begin{equation}\label{additive}
	f(E)=\sum_{i=1}^{|E|} \phi_f(E_{1, \dots,i})
\end{equation}
where $\phi_f$ is some function of events and $E_{1,\dots,i}$ denotes
the restriction of $E$ to its first $i$ elements. 
A simple example is the \emph{amount of lost edges} $M(E)$ that was
suffered as a result of $E$. Here, 
\begin{equation}\label{lost}
	\phi_M(E_{1,\dots, i})=\left\{
		\begin{array}{ll}
			2, \quad & E_i=F 
			\wedge D(E_{1,\dots, i-1})\neq \stale, \\
			0, & \text{else}.
		\end{array}
	\right.
\end{equation}
Let us refer to observables as in Eq.\ (\ref{additive}) as
\emph{additive random variables}. The following lemma states that when
evaluating expectation values of additive variables, only their
\emph{step-wise mean}
\begin{equation*}
	\bar\phi(E_{1\dots i}):= \big(\phi(E_{1,\dots,i-1}, S)
	+\phi(E_{1,\dots,i-1}, F)\big)/2
\end{equation*}
enters the calculation. 

\begin{lemma}[Expectation values of additive 
random variables]\label{lmMean}
	Let $f$ be an additive random variable. Set 
	\begin{equation*}
		\bar f(E):=\sum_{i=1}^{|E|} \bar\phi(E_{1\dots i}).
	\end{equation*}
	Then $\langle f \rangle = \langle \bar f \rangle$.
\end{lemma}

\begin{proof}
	Set $n=n_{\text T}$. We then have, by definition,
	\begin{eqnarray*}
		\langle f \rangle 
		&=& 2^{-n} 
		\sum_{E, |E|=n} \sum_{i=1}^n \phi(E_{1,\dots,i})  \\
		&=& \sum_{i=1}^n 2^{-i} \sum_{E, |E|=i} \phi(E)\\
		&=& \sum_{i=1}^n 2^{-i} \sum_{E, |E|=i} \bar\phi(E)
		= \langle \bar f \rangle.
	\end{eqnarray*}
\end{proof}

Note that
\begin{equation*}
	\bar \phi_M(E_{1,\dots, i})=
	\left\{ 
		\begin{array}{ll}
			1, \quad & D(E_{1,\dots, i-1})\neq \stale ,\\
			0, & \text{else},
		\end{array}
	\right.
\end{equation*}
in other words, $\bar\phi_M$ counts the \emph{number of attempted
fusions} $T$. Using Lemma \ref{lmMean}, we see that the expected
number of lost edges equals the expected number of fusion attempts:
$\langle M \rangle = \langle T \rangle$. This proves Lemma
\ref{lm:Attempts}.

In the following proof of  Theorem \ref{tm:Razor}, we will employ the identity
picture introduced in Section \ref{sc:Concepts}. The argument is broken
down into a series of lemmas.

\begin{lemma}[More is better than less]\label{lmMore}
	Let $I$ be a configuration. Then, for all $i$, 
	$Q(I+e_i) \geq Q(I)$.
\end{lemma}

\begin{proof}
	The proof is by induction on two parameters: on the 
	number of chains $|C|$
	and on the total length $L(C)$.
	To base the induction in both variables, we note that the claim is
	trivial if either $|C|\leq 1$ or $L\leq 2$.
	
	Now consider any configuration $C$. Let $S$ be the optimal strategy
	and denote by $C_S$ and $C_F$ the configurations created by $S(C)$
	in case of success and failure respectively. It is simple to check
	that $S(C)$ acting on $C+e_i$ yields $C_S+e_i$ or $C_F+e_i$. Hence 
	\begin{eqnarray*}
		Q(C+e_i)
		&\geq&1/2 \big(Q(C_S+e_i)+Q(C_F+e_i)\big).
	\end{eqnarray*}
	But unless $|C|\leq 1$ we have that in any event $E\in\{S,F\}$
	either $|C_{E}|<|C|$ or $L(C_{E})<L(C)$ and thus the
	claim follows by induction.
\end{proof}

\begin{lemma}[Winning is better than losing]\label{lmWin}
	Let $C\in \mathcal{C}$, let $C_S$ be the configuration resulting
	from the action of the optimal strategy on $C$ in the case of
	success, let $C_F$ be the obvious analogue. Then $Q(C_S)\geq
	Q(C_F)$.
\end{lemma}

\begin{proof}
	Let $\langle k,l \rangle$ be the action defined above. Clearly, 
	$C_F = C - e_k - e_l$. By the last lemma, $Q(C_F)\leq Q(C)$. But
	$Q(C)$ is the average of $Q(C_F)$ and $Q(C_S)$; 
	hence 
	\begin{equation*}
		Q(C_S)\geq
	Q(C)\geq Q(C_F).
	\end{equation*}
\end{proof}

\begin{lemma}[No catalysis]\label{lmCat}
	Let $C\in\mathcal{C}$. Then, for all $i$,
	$Q(C+e_i) \leq Q(C)+1$.
\end{lemma}

\begin{proof}
	We show the equivalent statement: for $C$ and $i$ s.\,t. $C_i\neq 0$
	it holds that $Q(C-e_i)\geq Q(C) -1$. Once more, the proof is by
	induction on $|C|, L$ and the validity of the claim for $|C|\leq 1$ or
	$L\leq 2$ is readily verified.

	Let $C, S, C_S, C_F$ be as in the proof of Lemma \ref{lmMore}.  If
	the application of $S(C)$ and the subtraction of $e_i$ commute, we
	can proceed as we did in Lemma \ref{lmMore}. A moment of thought
	reveals that this is always the case if not $C_i=1$ and
	$S(C)=\langle i, k \rangle$ (or, equivalently, $\langle k, i
	\rangle$) for some $k$. In fact, in this case we have
	\begin{eqnarray*}
		C_S&=&(\dots, l_i+l_k, \dots) \\
		C_F&=&(\dots, l_i-1, \dots, l_k-1, \dots),
	\end{eqnarray*}
	so that $C_F-e_i$ would take on a negative value at the $i$-th
	position. Note, however, that $C-e_i=C_S-e_i$. By induction it holds
	that $Q(C_S-e_i)\geq Q(C_S) - 1$ and further, by Lemma \ref{lmWin},
	$Q(C_S) - 1 \geq Q(C) -1$ which concludes the proof.
\end{proof}

\begin{lemma}[Fewer edges -- fewer fusions]\label{lmLessLess}
	Let $C\in \mathcal{C}, i$ be such that $C_i \neq 0$. Then
	\begin{equation*}
		\langle T \rangle(C - e_i) \leq \langle T \rangle(C),
	\end{equation*}
	where the expectation values are taken with respect to the respective
	optimal strategies.
\end{lemma}

\begin{proof}
	We will show that, for every $C \in \mathcal{C}$, the optimal
	strategy acting on $C':=C-e_i$ will content itself with a lower
	number of average fusion attempts $\langle T \rangle(C')$ than will
	the optimal strategy acting on $C$.
	Recall that Lemma~\ref{lm:Attempts} states
	\begin{eqnarray*}
		Q(C) &=& L(C) - \langle T \rangle(C). 
	\end{eqnarray*}
	Combining this and Lemma \ref{lmCat} we find
	\begin{eqnarray*}
		&&Q(C')\geq Q(C) - 1  \nonumber\\
		\Leftrightarrow&&
		L(C)-1 -\langle T \rangle(C') \geq L(C) -\langle T \rangle(C)-1 \nonumber \\
		\Leftrightarrow&&
		\langle T \rangle(C') \leq \langle T \rangle(C).
	\end{eqnarray*}
\end{proof}

We are finally in a position to tackle the original problem. 

\begin{proof}\emph{(of Theorem \ref{tm:Razor})}
	Let $C_\emptyset$ be some configuration. We will build a strategy
	which is valid on $C_\emptyset$ in the razor model and uses a fewer
	number of expected fusions than the optimal strategy in the full
	setup.  Define the shaving operator $\hat R: \mathcal{C} \to \mathcal{C}$
	which sets the length of each chain of length $i$ in the configuration it acts
	on to $\max(i,R)$. By a repeated application of the relation stated
	in Lemma \ref{lmLessLess}, we see that $\langle T \rangle(\hat R C) \leq
	\langle T \rangle(C)$.

	We build the razor model strategy's decision function $D'$
	inductively for all events in $\mathcal{E}_i$, for increasing $i$.
	Consider an event $E\in\mathcal{E}_i$. Denote by $C'_E$ the
	configuration resulting from $C_\emptyset$ under the action of $D'$
	in the event of $E$. $C'_E$ is well-defined as only the values of
	$D'$ for events with length smaller than $i$ enter its definition. 
	Set $D'(E)$ to the action taken by the optimal strategy for $\hat R C_E'$. 

	It is simple to verify that $D'$ defines a valid strategy for the
	razor model. By the results of the first paragraph, the expected
	number of fusions decreased in every step of the construction of
	$D'$. The claim follows.
\end{proof}

\subsection{An analytical bound --  random walk}
\begin{figure}
  \includegraphics{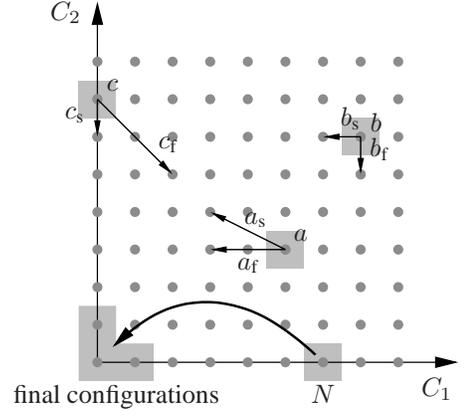}
  \caption{ The configuration space of the $R=2$ razor model is $\nn_0\times\nn_0$. Only the three actions $a$, $b$ and $c$ are available to reach the final configurations
    (exactly one EPR pair or GHZ state, or no chain at all), starting from the initial configuration that consists of $N$ EPR pairs. \label{fig:razor2} }
\end{figure}

Finally, we are in a position to prove an analytic upper bound on the
yield of any strategy building one-dimensional cluster chains. Quite
surprisingly, the description given by the razor model with a rather
radical parameter of $R=2$ is still faithful enough to deliver a good
bound as will be explained now.

In the $R=2$-model configurations are fully specified by giving the
number of EPR pairs $n_1$ and of chains of length two $n_2$ they
contain. Hence the configuration space is $\nn_0\times\nn_0$ and we
can picture it as the positive quadrant of a two-dimensional lattice.  
In each step a strategy can choose only among three non-trivial
actions: 
\begin{enumerate} 
	\renewcommand{\labelenumi}{(\alph{enumi})}
	\item Try to fuse two EPR pairs. We call this action $a$ for brevity.
	Let $C_S$ be the configuration resulting from a successful application
	of $a$ on $C$. Define the vector $a_S \in \zz \times \zz$ as $a_S :=
	C_S - C$. An analogous definition for $a_F$ and some seconds of
	thought yield 
	\begin{eqnarray*} 
		a_S&:=&(-2,1),\\ a_F&:=&(-2,0).
	\end{eqnarray*}
		
	\item
	Try to fuse two chains of length one and two, respectively. In the
	same manner as above we have
	\begin{eqnarray*}
		b_S &:=&(-1,0),\\
		b_F &:=&(0,-1).
	\end{eqnarray*}
	
	\item
	Try to fuse two chains of length two.
	\begin{eqnarray*}
		c_S &:=&(0,-1),\\
		c_F &:=&(2,-2).
	\end{eqnarray*}	
\end{enumerate}

The objective is to bound from below the minimum number of non-trivial
actions taken on average. Initially, we start with $N$ EPR pairs, so
$C_\emptyset=(N,0)$. As configuration space is a subspace of $\nn_0\times \nn_0$, we can 
describe the situation by a random walk in a plane. 

Any strategy will apply the rules $a,b,c$ until
one of the points $(0,0), (1,0), (0,1)$ is reached (illustrated in Fig.~\ref{fig:razor2}). 
Our proof will be
lead by the following idea: by applying one of the three non-trivial
actions to a configuration $C$, we will move ``on average'' by
\begin{eqnarray*}
	\bar a&:=&(a_S+a_F)/2=(-2,1/2),\\
	\bar b&:=&(-1/2,-1/2)\quad\text{or}\\
	\bar c&:=&(1,-3/2),
\end{eqnarray*}
respectively. The minimum number of expected fusion steps should then
be given by the minimum number of vectors from $\{\bar a, \bar b, \bar
c\}$ one has to combine to reach the origin starting from $(N,0)$. 
This procedure amounts to an interchange of two averages. The aim is 
to reach the origin or a point with distance one to it on average
as quickly as possible.

To make this intuition precise, set
\begin{equation*}
	\phi_\delta(E_{1,\dots, i}) : = 
	D(E_{1,\dots,i-1})_{E_i}.
\end{equation*}
Recall that $D(E_{1,\dots,i-1})$ is one of $\{a,b,c,\stale\}$.
Given the event $E$, $\phi_\delta(E)$ is the last action applied
to the configuration.
For any event $E=\{S,F\}^n$ we require that 
\begin{equation}\label{ineq1}
	\delta(E):=\sum_i \phi_\delta(E_{1, \dots, i})\leq (-N+1,1),
\end{equation}
which implies in particular that the same bound holds for $\langle
\delta \rangle$.
Define $a(E)$ to be the number of times the strategy will have decided
to apply rule ``a'' in the chain of events $\{ E_{1,\dots, i}\,|\,
i=1,\dots,|E| \}$ leading up to $E$. Formally 
\begin{equation*}
	\phi_a(E_{1,\dots, i})=
	\left\{
		\begin{array}{ll}
			1,\quad & D(E_{1, \dots, i-1})=a, \\
			0,&\text{else},
		\end{array}
	\right.
\end{equation*}
and $a(E)=\sum_{i=1}^{|E|} \phi_a(E_{1,\dots, i})$.
Further,
\begin{eqnarray*}
	\bar\phi_\delta(E_{1,\dots, i})=
	\phi_a(E_{1,\dots, i}) \bar a + \dots +
	\phi_c(E_{1,\dots, i}) \bar c,
\end{eqnarray*}
where $\phi_b, \phi_c$ are defined in the obvious way. It follows that
\begin{eqnarray}
	\langle \delta \rangle &=& \langle \bar \delta \rangle = 
	\langle a \rangle \bar a + 
	\langle b \rangle \bar b + 
	\langle c \rangle \bar c 
	\leq (-N+1,1), \label{const1}\\
	\langle T \rangle &=& 
	\langle a \rangle +
	\langle b \rangle +
	\langle c \rangle. \label{const2}
\end{eqnarray}

\subsection{An analytical bound --  convex optimization program}

Therefore, if $\langle T \rangle$ originates from a valid strategy it
is necessarily subject to the constraints put forward in Eqs.\
(\ref{const1},\ref{const2}). For each $N\in \nn$, 
a lower bound for the minimum expected number
of losses is thus given by a linear program, so a certain convex optimization
problem:  We define
\begin{equation*}
	B:=
	\left[
	\begin{array}{cc}
	-2 & 1/2 \\
	-1/2 & -1/2 \\
	1 & -3/2\\
	\end{array}
	\right].
\end{equation*}
Then, this lower bounds can be 
derived from the optimal solution of 
the linear program  given by  
\begin{eqnarray*}
	\text{minimize} && (1,1,1) x^T\\
	\text{subject to} && 
	x B  \leq (-N + 1,1) ,\nonumber\\
	&& x\geq 0,
\end{eqnarray*}
where the latter inequality is meant as a component-wise
positivity. This is a minimization over a vector 
$x\in \rr^3$. In this way, the performance of the razor model
is reduced to solving a family of convex optimization problems.
According to Lemma \ref{LP}, the solution of 
this linear program delivers the optimal objective value satisfying
\begin{equation*}
	\langle T \rangle = 4N/5 -2
\end{equation*}
for $N\geq 6$.

\begin{lemma}[Duality for linear program]\label{LP}
The optimal objective values 
of the family of 
linear programs
\begin{eqnarray*}
	\text{minimize} && (1,1,1) x^T\\
	\text{subject to} && 
	x B  \leq (-N + 1,1) ,\nonumber\\
	&& x\geq 0,\nonumber
\end{eqnarray*}
are given by
\begin{equation*}
	(1,1,1) x^T_{\text{opt}} = 
	\left\{
	\begin{array}{ll}
	0,& N =1,\\
	(N-1)/2, & N =2,...,5,\\
	(4 (N-1) -6 )/5   , &  N \geq 6.
	\end{array}
	\right.
\end{equation*}
\end{lemma}

\begin{proof}
This can be shown making use of
Lagrange duality
for linear programs. The dual to the above
problem, referred to as primal problem, 
is found to be 
\begin{eqnarray*}
	\text{maximize} && (N-1,-1) y^T\\
	\text{subject to} && 
	- y B^T  \leq (1,1,1) ,\nonumber\\
	&& y\geq 0.\nonumber
\end{eqnarray*}
This is a
maximization 
problem in $y\in \rr^2$, again a linear program 
(moreover, a
duality gap can never appear, i.\,e., the objective values of 
the optimal solutions of the primal and the dual problems are
identical). By finding -- for each $N$ -- a  
solution of the dual problem, 
which is assumed by the primal problem, 
we have hence proven
optimality of the respective solution. For all $N$, 
this family of solutions  
can be determined to be 
\begin{equation*}
	y = 
	\left\{
	\begin{array}{ll}
	(0,0),& N=1,\\
	(1/2, 0), & N=2,...,5,\\
	(4/5, 6/5) , &  N\geq 6.
	\end{array}
	\right.
\end{equation*}
It is straightforward to show that these are solutions of 
the dual problem, and that the respective objective
values are attained by appropriate solutions of the 
primal problem, e.\,g.
\begin{equation*}
	x = 
	\left\{
	\begin{array}{ll}
	(0,0,0),& N=1,\\
	((N-1)/2,0,0), & N=2,...,5,\\
	(2N/5,2(N/5-1),0) , &  N\geq 6.
	\end{array}
	\right.
\end{equation*}
The solutions yield the objective values
stated in the lemma.
\end{proof}

We subsequently highlight the consequence of 
this proof: we find the bound to the quality of the
globally optimal strategy: this shows that asymptotically
(for $p_{\text s}=1/2$)
at least five EPR pairs have to be invested on 
average
(see also the subsequent section) per single
gain of an edge in the linear cluster state.

\begin{corollary}[Upper bound to globally optimal strategy]
The quality of the optimal strategy for $N\ge6$ is bounded from above
by
\begin{equation*}
	Q(N) \leq N/5 + 2.
\end{equation*}
\end{corollary} 
This is one of the main results of this work.

\section{An inverse question}

Recall that so far we treated the problem 
``given some fixed number of
input pairs, how long a single chain can be obtained on average?''. It
is also legitimate to ask ``how many 
input pairs are needed to produce a
chain of some fixed length with (almost) unity probability of
success?''. After all, we might need just a specific length
for a given task. In the present section we establish 
that both questions are
\emph{asymptotically equivalent}, in the sense 
that bounds for either
problem imply bounds for the other one.

\begin{theorem}[Resources for given resulting length, upper bounds]
	Let $S$ be some strategy, let
	\begin{equation*}
		\tilde Q_S(N) \geq \alpha N + \beta
	\end{equation*}
	be a lower bound to its yield for some $\alpha, \beta \in
	\mathbbm{R}$ and all $N\ge N_0$. Choose an $\varepsilon>0$. Then there
	exists a strategy $S'$ such that, if $S'$ acts on $(1/\alpha +
	\varepsilon) L$ EPR pairs, it will output a single chain not
	shorter than $L$ with probability approaching unity as $L\to\infty$.
\end{theorem}

\begin{proof}
	Choose a number $b\in\mathbbm{N}$. Set $N=(1/\alpha+\varepsilon)L$. 
	There are arbitrary large $L$ such that $b$ divides $N$ and we will
	presently assume that $L$ has this property. We comment on the
	general case in the end.
	
	The strategy $S'$ proceeds in two stages, labeled
	I and II, to be analyzed in turn.
	Firstly, we divide the $N$ input pairs into $B=N/b$ blocks of size
	$b$ and let $S$ run on each of these blocks. 

	Denote by $N_i$ the random variable describing the final output
	length of the $i$-th block, $i=1,...,B$. 
	The $N_i$ are independent, identically
	distributed variables satisfying $\langle N_i \rangle \geq \alpha N
	+ \beta$. Set 
	$ N_{\text I} = \sum_{i=1}^B  N_i$ (the roman $\text I$
	signifies that we are dealing with the expected total length after
	the \emph{first} stage of $S'$). As the $N_i$ are independent, the
	variance of $ N_{\text I}$ equals $B \sigma^2$, where
	$\sigma^2<\infty$ is the variance of any of the $ N_i$.  By
	Chebychev's inequality we have
	\begin{eqnarray*}
		P\left[ |N_{\text I} - \langle N_{\text I} \rangle| \geq 
		B^{3/4}
		\right] &\leq& \text{Var}(N_{\text I}) B^{-3/2}\nonumber\\
		& =&
		\sigma^2\,B^{-1/2}.
	\end{eqnarray*}
	In other words, the relation $|N_{\text I} - \langle N_{\text I}
	\rangle | < B^{3/4}$ holds almost certainly if we let $L$ (and
	hence $B$) go to infinity for any fixed $b$. The same is true in
	particular for the weaker statement 
	\begin{eqnarray*}
		N_{\text I} \geq \langle N_{\text I} \rangle - B^{3/4}
		\geq B(\alpha b + \beta) - B^{3/4}.
	\end{eqnarray*}

	In the second stage II, 
	$S'$ builds up a single chain out of the $B$
	ones obtained before. Irrespective of how 
	$S'$ goes about in detail,
	the process will stop after exactly $B-1$ successful fusions. Now
	choose any $\delta>0$. We claim that 
	asymptotically no more than
	$(1+\delta)(B-1)$ failures will have occurred before the strategy
	terminates. Indeed, consider an event $E$ of length
	$2(1+\delta/2)(B-1)$. By the law of large numbers, 
	$E$ contains no fewer
	than $B-1$ successes and not more than 
	$(1+\delta)(B-1)$ failures, almost
	certainly as $B\to \infty$.  Hence the final output 
	length $N_{II}$
	fulfills
	\begin{eqnarray*}
		P\left[
		N_{\text{II}} > B(\alpha b + \beta) - 
		B^{3/4} - 2 (1+\delta) B
		\right] \to 1
	\end{eqnarray*}
	as $B\to\infty$. Plugging in the definitions of $B,N$, the r.\,h.\,s. of
	the estimate takes on the form
	\begin{eqnarray*}
		L + L\big(\varepsilon\alpha + \frac{1}{b}
		\,f_1(\alpha,\beta,\delta,\varepsilon)\big) - \big( \frac{L}{b}\,
		f_2(\alpha,\varepsilon)\big)^{3/4},
	\end{eqnarray*}
	where $f_1, f_2$ are some (not necessarily positive) functions of
	the constants.  By choosing the block length $b$ large enough, we
	can always make the second summand positive. For large enough $L$,
	the positive second term dominates the negative third one and hence
	$N_{\text{II}}> L$ almost certainly as $L\to\infty$.

	Lastly consider the case where $L$ is such that $b$ does not divide
	$N$. Choose $L\geq b/\varepsilon$. We can decompose $N=k b + r$ where
	$r<b$ and hence $r/L<b/L\leq\varepsilon$. Set $\varepsilon'=\varepsilon-r/L$.
	By construction $N'=(1/\alpha-\varepsilon')L$ divides $b$ and therefore
	already $N'<N$ input pairs are enough to build a chain of length $L$
	asymptotically with certainty.
\end{proof}

\begin{theorem}[Resources for given resulting length, lower bounds]
	Let
	\begin{equation*}
		Q(N)\leq \alpha N + \beta
	\end{equation*}
	be some upper bound to the optimal strategy's performance. Choose an
	$\varepsilon>0$. Then there exists no strategy $S'$ 
	such that, if $S'$
	acts on $(1/\alpha-\varepsilon)L$ EPR pairs, it will output a single
	chain not shorter than $L$ with probability approaching unity as
	$L\to\infty$.
\end{theorem}

\begin{proof}
	Assume there is such a strategy $S'$. Then
	\begin{equation*}
		\lim_{N\to\infty} \frac{\tilde Q_{S'}(N)}{N}\geq 
		(1/\alpha-\varepsilon)^{-1}
		>\alpha.
	\end{equation*}
	Hence $\tilde Q_{S'}(N)$ is eventually larger than 
	$Q(N)$, which is
	a contradiction.
\end{proof}

Suppose one aims to build a linear cluster state of length $N$.
Combining the results of the present section with the findings of
Sections \ref{Lower}, \ref{Upper} yields that the
goal is achievable with unit probability if more than $6.6 N$ EPR
pairs are available. Similarly, one will face a finite probability of
failure in case there are less than $5 N$ input chains.  Both
statements are valid asymptotically for large $N$.

\section{Two-dimensional cluster states}\label{2d}

\subsection{Preparation prescription}

We finally turn to the preparation of two-dimensional cluster states,
which are universal resources for quantum computation
\cite{Oneway,GS,LongGS}.  To build up a two-dimensional $n\times n$
cluster state clearly requires the consumption of $O(n^2)$ EPR pairs.
That this bound can actually be met constitutes the main result of
this section: this question had been open so far, with all known
schemes exhibiting a worse scaling. From our previous derivations, we
already know that length-$n$ linear cluster chains can be built
consuming $O(n)$ entangled pairs. Hence it suffices to prove that
linear chains with an accumulated length of $O(n^2)$ can be combined to
an $n\times n$-cluster. Consequently, for the constructions to come,
we will employ linear chains -- as opposed to EPR pairs -- as the
basic building blocks.

Again, to actually connect two chains to form a two-dimensional
structure, probabilistic gates from arbitrary architectures may be
utilized. The following claim will hold for gates that delete a
constant amount of edges from the participating chains on failure
(maybe unequal for the two chains), but not splitting them (no
$\sigma_z$ error outcome).  In case of success it shall create
cross-like structures, again deleting a certain amount of edges (see
Fig.~\ref{fig:fusion2}). In particular, the quadratic scaling as such
is not altered by a possibly small probability of success $p_{\text
s}<1/2$.

\begin{figure}
  \includegraphics{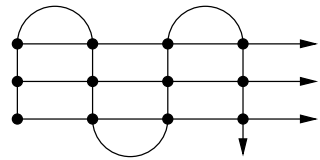}
  \caption{ A possible pattern of how to arrange
    $n+1$ linear clusters to build a two-dimensional cluster of width $n$.
    Fusion operations have to be applied at the black circles
    along the long linear cluster state.
    Free ends carrying spare overhead are shown as arrows.
  \label{fig:2d_strategy} }
\end{figure}

The main problem faced is to find a preparation scheme that does not
'tear apart' successfully prepared intermediate states in case of a
failed fusion. The challenge will be met by ({\it a}) switching from
type-I to type-II fusion (Section \ref{sc:typeii}) and ({\it b})
employing the pattern shown in Fig.~\ref{fig:2d_strategy} (Section
\ref{sc:2dproof}).

\subsection{Linear optical type-II fusion gate}
\label{sc:typeii}

As for linear optics fusion gates, an error outcome in the
type-I gate would tear each chain apart
where we tried to fuse. Hence the related type-II fusion
gate~\cite{terry} with a more suitable error outcome will be used.
How this one actually acts is shown in Fig.~\ref{fig:fusion2}.

In preparation of a fusion attempt, a ``redundantly encoded'' qubit with
two photons (see~\cite{terry}) is produced in one chain by a
$\sigma_x$ measurement, which consumes two edges (giving rise to
another $2n^2$ edges). Now the fusion type-II gate creates a
two-dimensional cross-like structure on success when being applied to
one of the photons in the redundantly encoded qubit and one of the
other chain's qubits.  In case of failure it acts like a $\sigma_x$
measurement, therefore decreasing the encoding level of the redundancy
encoded qubit by one and deleting two edges from the other chain,
leaving us with a redundancy encoded qubit there. Hence, we may apply
the fusion type-II again without any further preparation, deleting two
edges on successive failures from the two chains alternatingly.  For
convenience we assume that we lose two edges per involved chain per
failure instead. This increases the overhead requirement roughly by a
factor of two but allows us to forget about the asymmetry in the
fusion process. Hence, in the following any resource requirements will
be given in terms of double edges instead of single ones.

Similar to the type-I case, the optimal success probability can
be found. Actually this type of fusion gate should perform a Bell
state measurement, hence $p_{\text s}\le1/2$~\cite{CL01}.
In fact, the gate proposed in Ref.~\cite{terry} consists 
of the parity check, the Hadamard rotation and measurement of
the second qubit (see Fig.~\ref{fig:parity_bell}) with two additional
Hadamard gates applied before (which only map Bell states onto
Bell states).

\subsection{Asymptotic resource consumption for near-deterministic
cluster state preparation}
\label{sc:2dproof}

\begin{theorem}[Quadratic scaling of resource overhead]\label{lemma:weaving}
For any success probability $p_{\text s}\in(0,1]$ of type-II fusion,
an $n \times n$ cluster state can be prepared
using $O(n^2)$ edges
in a way such that the overall
probability of success approaches unity
\begin{equation*}
        P_{\text s}(n)\rightarrow 1
\end{equation*}
as $n\rightarrow \infty$.
 \end{theorem}

\begin{proof}
The aim is to prepare an $n\times n$ cluster state, starting from
$n+1$ one-dimensional chains. For any integer $l$,
starting point is a collection of
$n$ one-dimensional chains of length
$m=n+l$, and a single longer chain of
length $L = n (l+1)$, referred to subsequently
as thread. In order to achieve the goal,
a suitable choice for a pattern of
fusion attempts is required.
One such suitable ``weaving pattern'' is depicted in
Fig.~\ref{fig:2d_strategy}. Here, solid
lines depict linear chains, whereas dots
represent the vertices along the thread
where fusion gates are being applied.

The aim will then
be to identify a function $n\mapsto g(n)$
such that the choice $m=g(n) $ leads to the appropriate
scaling of the resources. In fact, it will turn out that a linear
function is already suitable, so for an $a>1/p_{\text s}$ we will
consider $g(n)=an$. This number
\begin{equation*}
        m-n= g(n)-n= (a-1)n 
\end{equation*}
quantifies the resource overhead: in
case of failure, one can make use
of this overhead to continue with the prescription
without destroying the cluster state. If this overhead is too
large, we fail to meet the strict requirements on the scaling
of the overall resource consumption, if it is too small,
the probability of failure becomes too large. Note that there
is an additional overhead reflected by the choice
$L$. This, however, is suitably chosen not to
have an implication on the asymptotic scaling of the
resources.

\begin{figure}
  \includegraphics{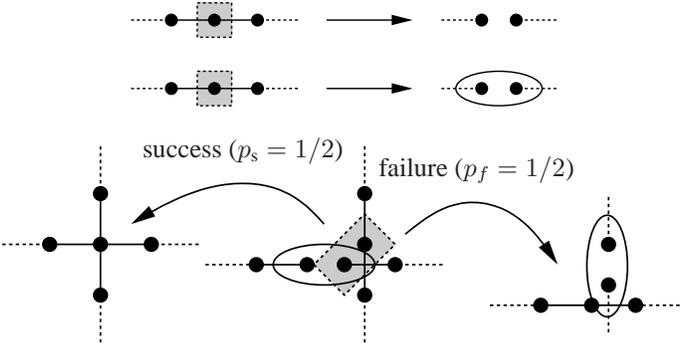}
  \caption{ The elementary linear optics tools for building two-dimensional structures from linear cluster chains. From top to bottom: a $\sigma_z$ measurement to
    remove unneeded nodes, a $\sigma_x$ measurement to create a
		redundancy encoded qubit in preparation of type-II fusion. The
		last figure shows the action
    of a fusion type-II attempt. \label{fig:fusion2} }
\end{figure}

Given the above prescription,
depending on $n$, the overall probability
$P_{\text s}(n)$ of succeeding to prepare an $n\times n$
cluster state can be written as
\begin{equation*} 
  P_{\text s}(n) = \pi_{\text s}(n)^n .
\end{equation*}
Here,
\begin{equation*}
  \pi_{\text s}(n) = p_{\text s}^n\sum\limits_{k=0}^{(a-1)n}
  (1-p_{\text s})^k{n+k-1 \choose k}
\end{equation*}
is the success probability to weave a single chain of length $an$ into the carpet of size $n$
with the binomial quantifying the number
of ways to distribute $k$ failures on $n$ nodes \cite{stanley}.
$p_{\text s}>0$ and $1-p_{\text s}$ are
the success and failure probabilities for a fusion attempt, respectively.
It can be rephrased as the probability to find at least $n$ successful outcomes in $an$ trials,
\begin{eqnarray*} 
        \pi_{\text s}(n) &=&
        \sum\limits_{k=n}^{a n} (1-p_{\text s})^{an-k}p_{\text s}^k { a n \choose k} 
        \nonumber\\
        &=&
        1- F(n-1, a n ,p_{\text s}).
\end{eqnarray*}
Here, $F$ denotes the standard cumulative
distribution function of the binomial distribution \cite{Cum}.
Since $n-1 \leq a n p_{\text s}$ for all $n$, as $a>1/p_{\text s}$ is assumed,
we can hence bound $\pi_{\text s}(n)$ from below
by means of {\it Hoeffding's inequality}
\cite{Hoeff,Hoeff2}, providing an exponentially decaying
upper bound of the tails of the cumulative
distribution function. This gives
rise to the lower bound
\begin{eqnarray*}
        \pi_{\text s}(n)\geq 1- \exp\left(
        -\frac{2 (a n p_{\text s} - n + 1)^2 }{a n}
        \right).
\end{eqnarray*}
Now, again since $a>1/p_{\text s}$, we have that
\begin{equation*}
  \pi_{\text s}^n \ge \left( 1-\exp(-cn) \right)^n
\end{equation*}
with $c:=2(ap_{\text s}-1)/a>0$. Further, for any $k\in\nn$
there exists an $n_0\in\nn$ such that for all $n\ge n_0$
\begin{equation*}
  \left( 1-\exp(-cn) \right)^n > \left(1-1/(kn)\right)^n.
\end{equation*}
Noticing
\begin{equation*} 
  \lim_{n\rightarrow\infty}\left(1-1/(kn)\right)^n = \me^{-1/k}
\end{equation*}
we can find for any $\varepsilon>0$
a $k$ satisfying $1-\me^{-1/k}<\varepsilon$.
Therefore, for any $\varepsilon>0$ it holds that $\lim_{n\rightarrow\infty}P_{\text s}>1-\varepsilon$.
This ends the argument leading to the appropriate
scaling.
\end{proof}

Even within the setting of quadratic resources,
the appropriate choice for $a$ does have an impact:
If the probability of success $p_{\text s}$ is
too small for a given $a$,
\begin{equation*}
        1/p_{\text s}>a>1, 
\end{equation*}
then this will
lead to $\lim_{n\rightarrow\infty} P_{\text s}(n) =0$,
so the preparation of the cluster
will eventually fail, asymptotically with certainty.
This sudden change of the asymptotic behavior
of the resource requirements, leading essentially to either
almost unit (almost all cluster states can successfully be prepared)
or almost vanishing success probability is a simple
threshold phenomenon as in {\it percolation theory}.
In turn, for a given $a$, $p_{\text{th}}=1/a$ can be
taken as a threshold probability: above this threshold
almost all preparations will succeed, below it they will fail
\cite{Percolation}. This number $a$ essentially
dictates the constant factor in front of the quadratic
behavior in the
scaling of the resource requirements. Needless to
say, this depends on $p_{\text s}$.

This analysis shows that a two-dimensional
cluster state can indeed be prepared using $O(n^2)$ edges,
employing probabilistic quantum gates only. This can be
viewed as good news, as it shows that the natural scaling of
the use of such resources can indeed be met, with asymptotically
negligible error. Previously, only strategies leading to a
super-quadratic resource consumption have been known.
In turn, any such other scaling of the resources
could have been viewed as a threat to the possibility of
being able to prepare higher-dimensional cluster states
using probabilistic quantum gates.

\section{Summary, discussion, and outlook}

In this paper, we have addressed the question of 
how to prepare cluster states using probabilistic gates.
The emphasis was put on finding bounds that the optimal
strategy necessarily has to satisfy, to identify final bounds
on the resource overhead necessary in such a preparation.
This issue is particularly relevant in the context of
linear optics, where the necessary overhead in resource
is one of the major challenges inherent in this type
of architecture. 
It turns out that the way the classical strategy is chosen
has a major impact on the resource consumption. By 
providing these rigorous bounds, we hope to give a 
guideline to the feasibility of probabilistic state generation.
{\it One central observation, e.\,g., is that for any preparation
of linear cluster states using linear optical gates as
specified above, one necessarily needs at least
five EPR pairs per average gain of one edge. }
This limit can within these rules no further be 
undercut. But needless to say, the derived results
are also applicable to other architectures, and we 
tried to separate the general statements from those
that focus specifically on linear optical setups.

It is also the hope that the introduced 
tools and ideas are applicable beyond the 
exact context discussed in the present paper.
There are good reasons to believe that these 
methods may prove useful even
when changing the rules: For example, 
as fusion type-II can also
be used for production of redundancy encoding resource
states~\cite{Ralph05} and linear cluster states 
in a similar fashion,
similar bounds to
resource consumption may be derived for these schemes. Due
to the fact that fusion type-II does not require
photon number resolving detectors, this could be a matter
of particular interest for experimental realizations.
Also, generalizations of some of the statements for $p_{\text s}\ne1/2$
have been explicitly derived. 
Other generalizations may well also be proven
with the tools developed in this paper.

Concerning lossy operations, we emphasize again
that when all EPR pairs are simultaneously
created in the beginning, their storage time will be
minimized by application of the strategy that optimizes
the expected final length. Obviously, 
the problem of storage using
fiber loops or memories is a key issue in any
realization. Yet, for a given loss mechanism, 
it would be interesting to see to what extent a
modification of the optimal protocol would follow
-- compared to the one here assuming perfect
operations --  depending on the figures of merit 
chosen. One
then expects trade-offs between different
desiderata to become relevant \cite{Rohde}.
In the way it is done here, decoherence induced
by the actual gates employed for the fusion process
is also minimized exactly 
by choosing the optimal strategy of this work:
it needs the least number of uses 
of the underlying quantum gates.

Further, studies in the field of fault tolerance may well
benefit from this approach. To start with, one has to be aware
that the overhead induced in fully fault-tolerant one-way
computing schemes is quite enormous \cite{Trees}.
This is extenuated when considering photon loss
only as a source of errors \cite{Trees2,Ralph05}. 
Yet, methods
as the ones presented here will be expected to be
useful to very significantly reduce the number of 
gate invocations in the preparation of the resources.

\section{Acknowledgments}

We would like to acknowledge discussions with the participants 
of the LoQuIP conference on linear optical quantum 
information 
processing in
Baton Rouge in April 2006, organized by J.\ Dowling.
The authors are grateful to A.\ Feito for a
host of comments on a draft version. This work has 
been supported by
the DFG, the EU (QAP), the EPSRC, QIP IRC, Microsoft  
Research through the European PhD Scholarship Programme, and the EURYI
award scheme.


\begin{thebibliography}{99}
\bibitem{KLM}
        E.\ Knill, R.\ Laflamme, and G.J.\ Milburn,
        Nature (London)
        \textbf{409}, 46 (2001).

\bibitem{Rev}
         P.\ Kok, W.J.\ Munro, K.\
         Nemoto, T.C.\ Ralph, J.P.\ Dowling, and G.J.\ Milburn,
         Rev.\ Mod.\ Phys.\ \textbf{79}, 135 (2007);
         C.R.\ Myers and R.\ Laflamme, quant-ph/0512104.

\bibitem{Reznik}
        N.\ Yoran and B.\ Reznik,
        Phys.\ Rev.\ Lett.\ \textbf{91}, 037903 (2003).

\bibitem{Nielsen04}
          M.A.~Nielsen,
        Phys.\ Rev.\ Lett.\  \textbf{93}, 040503 (2004).

\bibitem{terry}
         D.E.\ Browne and T.\ Rudolph,
         Phys.\ Rev.\ Lett.\ {\bf 95}, 010501 (2005).

\bibitem{Ralph05}
        T.C.\ Ralph, A.J.F.\ Hayes, and A.\ Gilchrist,
        Phys.\ Rev.\ Lett. {\bf 95}, 100501 (2005);
        A.\ Gilchrist, A.J.F.\ Hayes, T.C.\ Ralph, quant-ph/0505125.
        
\bibitem{PC}
        T.B.\ Pittman, B.C.\ Jacobs, and J.D.\ Franson,
        Phys.\ Rev.\ A {\bf 64}, 062311 (2001).

\bibitem{Dist}
        J.I.\ Cirac, A.\ Ekert, S.F.\ Huelga, and C.\
        Macchiavello, Phys.\ Rev.\ A {\bf 59}, 4249 (1999).

\bibitem{Dist2}
        J.\ Eisert, K.\ Jacobs, P.\ Papadopoulos, and
        M.B.\ Plenio, Phys.\ Rev.\ A {\bf 62}, 052317 (2000);
        D.\ Collins, N.\ Linden, and S.\ Popescu,
        \emph{ibid.}\ {\bf 64}, 032302 (2001).

\bibitem{Computers}
	M.A.\ Nielsen and I.L.\ Chuang,
	{\it Quantum computation and quantum
	information} (Cambridge University Press,
	Cambridge, 2000); J.\ Eisert and M.M.\
	Wolf, {\it Quantum computing}, in 
	{\it Handbook of nature-inspired and innovative 	
	computing} (Springer, New York, 2006).

\bibitem{SP}
        J.\ Eisert, Phys. Rev. Lett. \textbf{95}, 040502 (2005);
        S.\ Scheel,  W.J.\ Munro, J.\ Eisert, K.\ Nemoto,
        and P.\ Kok, Phys.\ Rev.\ A {\bf 73}, 034301 (2006);
         S.\ Scheel and K.M.R.\ Audenaert,
         New J.\ Phys.\ {\bf 7}, 149 (2005);
         S.\ Scheel and N.\ L{\"u}tkenhaus,
        \emph{ibid.}\ \textbf{6}, 51 (2004);
        E.\ Knill, Phys.\ Rev.\ A \textbf{68}, 064303 (2003).

\bibitem{Oneway}
        H.-J.\ Briegel and R.\ Raussendorf,
        Phys.\ Rev.\ Lett.\ {\bf 86}, 910 (2001);
        R.\ Raussendorf and H.-J.\ Briegel,
        \emph{ibid.}\ {\bf 86},  5188 (2001).

\bibitem{GS}
        M.\ Hein, J.\ Eisert, and H.-J.\ Briegel,
        Phys.\ Rev.\ A {\bf 69}, 062311 (2004);
         M.\ Van den Nest, J.\ Dehaene, and B.\ De Moor,
         \emph{ibid.}\ {\bf 69}, 022316 (2004);
        R.\ Raussendorf, D.E.\ Browne, and H.-J.\ Briegel,
        \emph{ibid.}\ {\bf 68}, 022312 (2003);
        D.\ Schlingemann and R.F.\ Werner,
        \emph{ibid.}\ {\bf 65}, 012308 (2002).

\bibitem{LongGS}
        M.\ Hein, W.\ D{\"u}r, J.\ Eisert,
        R.\ Raussendorf, M.\ Van den Nest, and
        H.-J.\ Briegel,
        quant-ph/0602096;
        D.E.\ Browne and H.-J.\ Briegel, quant-ph/0603226.

\bibitem{Exp}
         P.\ Walther, K.J.\ Resch, T.\ Rudolph, E.\ Schenck,
         H.\ Weinfurter, V.\ Vedral, M.\ Aspelmeyer, and A.\
         Zeilinger, Nature {\bf 434}, 169 (2005);
         N.\ Kiesel, C.\ Schmid, U.\ Weber, G.\ Toth, O.\ G\"uhne,
         R.\ Ursin, and H.\ Weinfurter, Phys.\ Rev.\ Lett.\ 
         {\bf 95}, 210502 (2005).

\bibitem{EPR}
	The term ``EPR pair'' is to be understood in the sense of dual
	rail encoding introduced in Section \ref{sc:FusionGates}.

\bibitem{BK}
         S.D.\ Barrett and P.\ Kok,
         Phys.\ Rev.\ A {\bf 71}, 060310(R) (2005);
         Y.L.\ Lim, S.D.\ Barrett, A.\ Beige, P.\ Kok, and 
         L.C.\ Kwek,
    	Phys.\ Rev.\ A {\bf 73}, 012304 (2006).

\bibitem{O2}        
         S.C.\ Benjamin, Phys.\ Rev.\ A {\bf 72}, 056302 (2005);
          Q.\ Chen, J.\ Cheng, K.-L.\ Wang, and J.\ Du,
          \emph{ibid.}\ {\bf 73}, 012303 (2006);
          G.\ Gilbert, M.\ Hamrick, and
          Y.S.\ Weinstein, \emph{ibid.} {\bf 73}, 064303 (2006).

\bibitem{DR}
        L.M.\ Duan and R.\ Raussendorf,
        Phys.\ Rev.\ Lett.\ {\bf 95}, 080503 (2005).

\bibitem{Letter}
        K.\ Kieling, D.\ Gross, and J.\ Eisert,
        J. Opt. Soc. Am. B {\bf 24}, 184 (2007).
        
\bibitem{Trees}
        C.M.\ Dawson, H.L.\ Haselgrove, and M.A.\ Nielsen,
        Phys.\ Rev.\ Lett.\ {\bf 96}, 020501 (2006);
        C.M.\ Dawson, H.L.\ Haselgrove, and M.A.\ Nielsen,
        Phys.\ Rev.\ A\ {\bf 73}, 052306 (2006);
        P.\ Aliferis and D.W.\ Leung,
        Phys.\ Rev.\ A {\bf 73}, 032308 (2006);
        R.\ Raussendorf, J.\ Harrington, and K.\ Goyal,
        quant-ph/0510135.

\bibitem{Trees2}
         M.\ Varnava, D.E.\ Browne, and T.\ Rudolph,
        Phys.\ Rev.\ Lett.\ {\bf 97}, 120501 (2006).
                
\bibitem{Port}
      S.C.\ Benjamin, J.\ Eisert, and T.M.\ Stace,
      New J.\ Phys.\ {\bf 7}, 194 (2005).

\bibitem{Plenio}
	S.\ Bose, P.L.\ Knight, M.B.\ Plenio, and V.\ Vedral,
	Phys.\ Rev.\ Lett.  {\bf 83}, 5158 (1999);
	D.E.\ Browne, M.B.\ Plenio, and S.F.\ Huelga,
	\emph{ibid.}\ {\bf 91}, 067901 (2003).

\bibitem{VK}
        N.G.\ Van Kampen, {\it Stochastic processes in 
        physics and chemistry} (North Holland, Amsterdam, 1992).

\bibitem{Convex}
    S.\ Boyd and L.\ Vandenberghe, {\it Convex optimization}
    (Cambridge University Press, Cambridge, 2004).


\bibitem{Chris}
	Compare also ``An EPR pair is nothing but a cluster
	state of length one'', C.M.\ Dawson, LoQUIP, Baton
	Rouge, April 2006.

\bibitem{CL01}
    J.~Calsamiglia and N.~L{\"u}tkenhaus, 
    Appl. Phys. B {\bf 72}, 67--71 (2001).

\bibitem{sloane}
    N.J.A.\ Sloane,
    {http://www.research.att.com/projects/OEIS? Anum=A000070}.

\bibitem{Kieling05a}
        K.\ Kieling, {\it Linear optical methods in quantum information
        processing} (Diploma thesis, University of Potsdam, 2005).


\bibitem{twoChainsGeneral}
    With $p_{\text s}\ne1/2$ one obtains
    \begin{eqnarray*}
      Q(C) 
       &\ge& l_1+l_2-2(1-p_{\text s})/p_{\text s}.
    \end{eqnarray*}

\bibitem{kChainsGeneral}
    In case of $p_{\text s}\ne1/2$
    \begin{equation*}
      \tilde Q\left(\sum_iC_{(i)}\right) \ge \sum_i\tilde Q_{S'}(C_{(i)})-2(1-p_{\text s})/p_{\text s}(k-1)
    \end{equation*}
    holds.

\bibitem{table}
        For the full table, see
        www.imperial.ac.uk/quantuminformation.

\bibitem{stanley}
    R.P.~Stanley, {\it Enumerative combinatorics} 
    (Wadsworth \& Brooks, 1986).

       
\bibitem{Cum}
    That is, 
    \begin{equation*}
      F(k,n,p) = \sum_{l=0}^k {n \choose l}
      p^l (1-p)^{n-l}.
    \end{equation*}

\bibitem{Hoeff}
    W.\ Hoeffding, J.\ Am.\ Stat.\ Ass.\ {\bf 58}, 13 (1963).

\bibitem{Hoeff2}
    Hoeffding's inequality states that 
    \begin{equation*}
    	F(k,n,p)\leq \exp(-2 (np-k)^2/n) 
    \end{equation*}	 
    for $k<n p$.

\bibitem{Percolation}
        G.\ Grimmet, {\it Percolation} (Springer, New York, 1999).

\bibitem{Rohde}
        P.P.\ Rohde, T.C.\ Ralph, and W.J.\ Munro,
        Phys.\ Rev.\ A\ \textbf{75}, 010302(R) (2007).

\end{thebibliography}
\end{document}